\pdfoutput=1
\documentclass{siamltex1213}[11pt]
\usepackage{amsmath,graphicx,amssymb,tikz,microtype}
\usetikzlibrary{shapes,arrows}
\usetikzlibrary{positioning}
\renewcommand{\vec}[1]{\mathbf{#1}}
\newcommand{\vq}{\vec{q}}
\newcommand{\vF}{\vec{F}}
\newcommand{\ti}{\text{in}}
\newcommand{\te}{\text{ext}}
\newcommand{\bp}{\begin{pmatrix}}
\newcommand{\ep}{\end{pmatrix}}

\usepackage{hyperref}

\definecolor{webgreen}{rgb}{0,.35,0}
\definecolor{webbrown}{rgb}{.6,0,0}
\definecolor{RoyalBlue}{rgb}{0,0,0.9}
\definecolor{purp}{rgb}{0.6,0.05,0.8}
\definecolor{ora}{rgb}{0.7,0.35,0.02}

\hypersetup{
   colorlinks=true, linktocpage=true, pdfstartpage=3, pdfstartview=FitV,
   breaklinks=true, pdfpagemode=UseNone, pageanchor=true, pdfpagemode=UseOutlines,
   plainpages=false, bookmarksnumbered, bookmarksopen=true, bookmarksopenlevel=1,
   hypertexnames=true, pdfhighlight=/O,
   urlcolor=webbrown, linkcolor=RoyalBlue, citecolor=webgreen,
   pdfauthor={Anna Lieb, Chris H. Rycroft, and Jon Wilkening},
   pdfsubject={Optimizing intermittent water supply in pipe distribution networks},
   pdfkeywords={},
   pdfcreator={pdfLaTeX},
   pdfproducer={LaTeX with hyperref}
}

\title{Optimizing intermittent water supply in urban pipe distribution
networks\thanks{This work was supported by the Director, Office of Science,
Computational and Technology Research, U.S. Department of Energy under contract
number DE-AC02-05CH11231, and by the National Science Foundation Graduate Research Fellowship Program under Grant No. DGE 1106400.}}
\author{Anna M.~Lieb\footnotemark[2]\ \footnotemark[3]
\and Chris H.~Rycroft\footnotemark[4]\ \footnotemark[3]
\and Jon Wilkening\footnotemark[2]\ \footnotemark[3]}

\begin{document}
\maketitle
\renewcommand{\thefootnote}{\fnsymbol{footnote}}
\footnotetext[2]{Department of Mathematics, University of California, Berkeley, CA 94720.}
\footnotetext[3]{Department of Mathematics, Lawrence Berkeley Laboratory, Berkeley, CA 94720.}
\footnotetext[4]{Paulson School of Engineering and Applied Sciences, Harvard University, MA 02138.}
\renewcommand{\thefootnote}{\arabic{footnote}}

\begin{abstract}
  In many urban areas of the developing world, piped water is supplied only
  intermittently, as valves direct water to different parts of the water
  distribution system at different times. The flow is transient, and may
  transition between free-surface and pressurized, resulting in complex
  dynamical features with important consequences for water suppliers and users.
  Here, we develop a computational model of transition, transient pipe flow in
  a network, accounting for a wide variety of realistic boundary conditions. We
  validate the model against several published data sets, and demonstrate its
  use on a real pipe network. The model is extended to consider several
  optimization problems motivated by realistic scenarios. We demonstrate how to
  infer water flow in a small pipe network from a single pressure sensor, and
  show how to control water inflow to minimize damaging pressure transients.
\end{abstract}

\section{Introduction}
From the dry taps of Mumbai to the dusty reservoirs of S\~ao Paolo, urban water scarcity is a common condition of the present, and a likely feature of the future. Hundreds of millions of people worldwide are connected to water distribution systems subject to intermittency. This intermittent water supply may take many forms, from unexpected disruptions to planned supply cycles where pipes are filled and emptied regularly to shift water between different parts of the network at different times~\cite{kumpel2013,Vairavamoorthy2008}. In Mumbai, for example, Vaivaramoorthy~\cite{Vairavamoorthy2008} reports that on average, residents have water flowing from their taps less than 8 out of 24 hours. Intermittent supply is often inequitable, with low-income neighborhoods experiencing lower water pressure and shorter supply durations than high-income ones~\cite{Sharma2009}. Intermittent supply not only limits water availability, but also compromises water quality and damages infrastructure. With field data from urban India, Kumpel and Nelson~\cite{Kumpel2014} quantified the deleterious effect of intermittency on water quality, showing that both the initial flushing of water through empty pipes---as well as periods of low pressure---corresponded with periods of increased turbidity and bacterial contamination. Christodoulou~\cite{Christodoulou2012} observed when that a drought in Cypress ushered in two years of intermittent supply, pipe ruptures increased by 30\%--70\% per year.

Whereas intermittent water supply creates challenges for water managers and water users, the phenomenon creates opportunities for applied mathematics. It is an interesting and difficult mathematical problem to efficiently model transient pipe flow in networks---including transitions to and from pressurized states---with uncertain or complex boundary conditions. In this work we introduce a framework to not only describe intermittent water supply, but also use optimization to improve either our description of the system, or the operation of the described system in order to reduce risks such as infrastructure damage.

Intermittent supply falls in somewhat of a modeling gap. Water distribution software abounds, including the free and open source software EPANET~\cite{EPA} produced by the US government, as well as many commercial packages~\cite{Ghidauoi2005}. Yet, to the authors' knowledge, all these fail to account for filling, emptying, and instances of subatmospheric pressure---phenomena that are vitally important for users and managers dealing with intermittent supply. Sewer system software such as the Storm Water Management Model (SWMM)~\cite{SWMM} and Illinois Transient Model (ITM)~\cite{ITM} to some extent handle the physics of interest, but are packaged in elaborate graphical user interfaces and are not readily amenable to model improvements or optimization.

Furthermore, the authors have encountered a relative paucity of research work dealing with modeling intermittent supply. The work of De Marchis~\cite{DeMarchis2010} explicitly studies filling and emptying in a water distribution system in Palermo, Italy, but with a method of characteristics implementation of the classical water hammer equations. This treatment assumes pipes are either entirely dry or entirely full, and that air pressure inside the pipes is always atmospheric. After calibrating a friction parameter, they found about 5\% agreement with empirical data. Subsequent work reported by De Marchis~\cite{DeMarchis2013} uses the same model to assess losses in the distribution system. Freni~\cite{Freni2014}, uses this model to determine pressure valve settings to reduce distribution inequality, but through scenario comparison rather than optimization. For sewer flow, Sanders~\cite{Sanders2011} presents a network implementation of the two-component pressure approach (TPA) of Vasconcelos~\cite{Vasconcelos2006}. The modeling for ITM was published by Le\'on in~\cite{Leon2010}. Urban water drainage is coupled to free surface flow by Borsche and Klaar~\cite{Borsche2014}. Note that Buosso et al.~\cite{Bousso2013} give a general review of the storm water drainage literature with more details than we have provided here.

The present work comprises an effort to address the scarcity of tools available for those interested in modeling the details of intermittent supply, and to specifically incorporate such tools within an optimization framework. We use an underlying model of coupled systems of one-dimensional hyperbolic conservation laws that strikes a balance between real-world relevance and both computational and theoretical tractability. Our computational framework will allow for straightforward implementation of alternative physical models in future studies.

\section{Model}
The Preissman slot formulation~\cite{Preissman1961} is used to describe flow within each pipe, building on existing literature for transient, transition flow in closed conduits. The flow is assumed to be inviscid and incompressible. The dynamical description considers depth-averaged flow within a modified geometry that permits a single set of equations to describe both free-surface and pressurized flow. Consideration of one-dimensional dynamics is a reasonable approximation given that the ratio of pipe diameter $D$ to pipe length $L$ is 1\% or smaller in realistic scenarios. 

\begin{figure}
  \begin{center}
    \begin{tabular}{cc}
      \includegraphics[width = 0.465\textwidth]{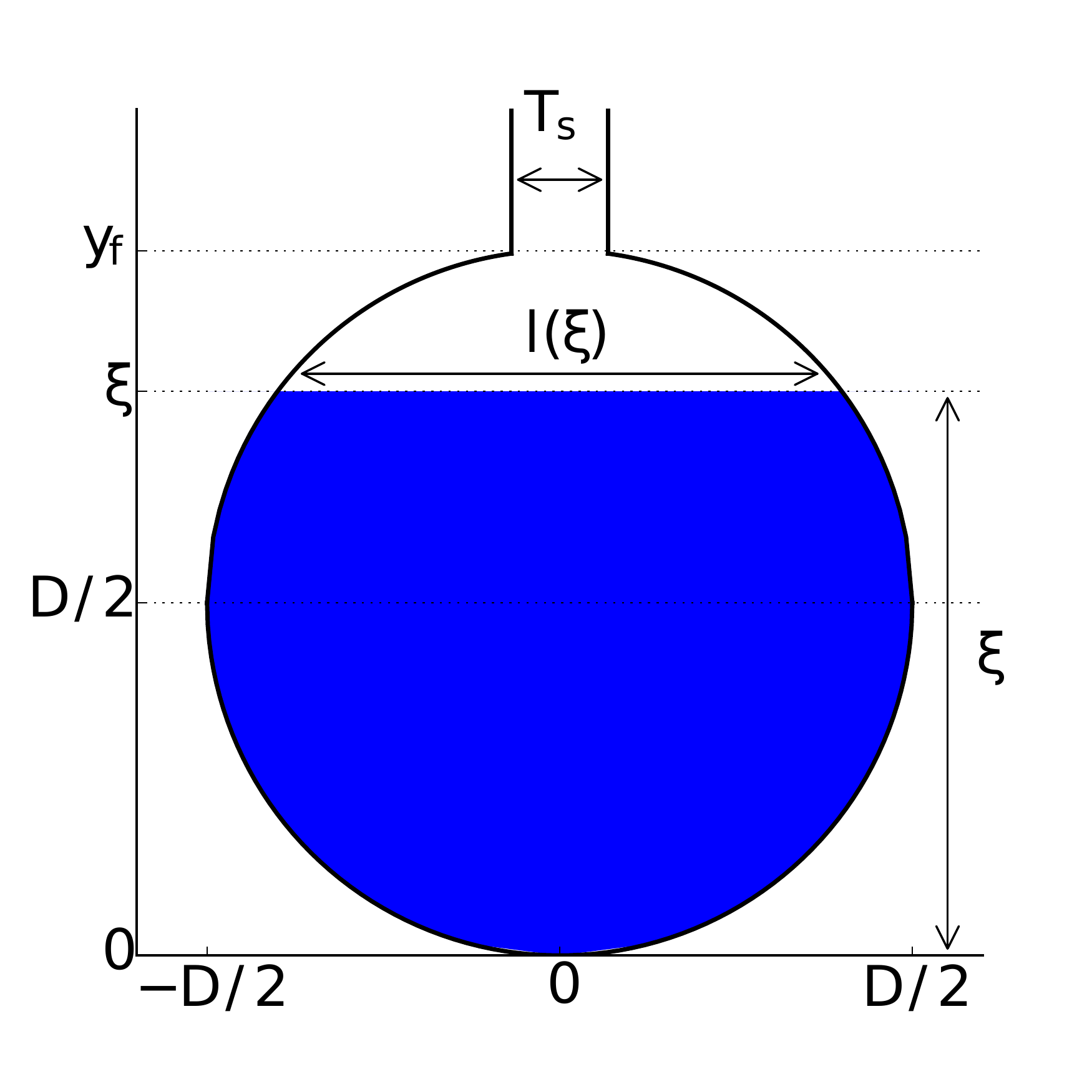} &
      \includegraphics[width = 0.465\textwidth]{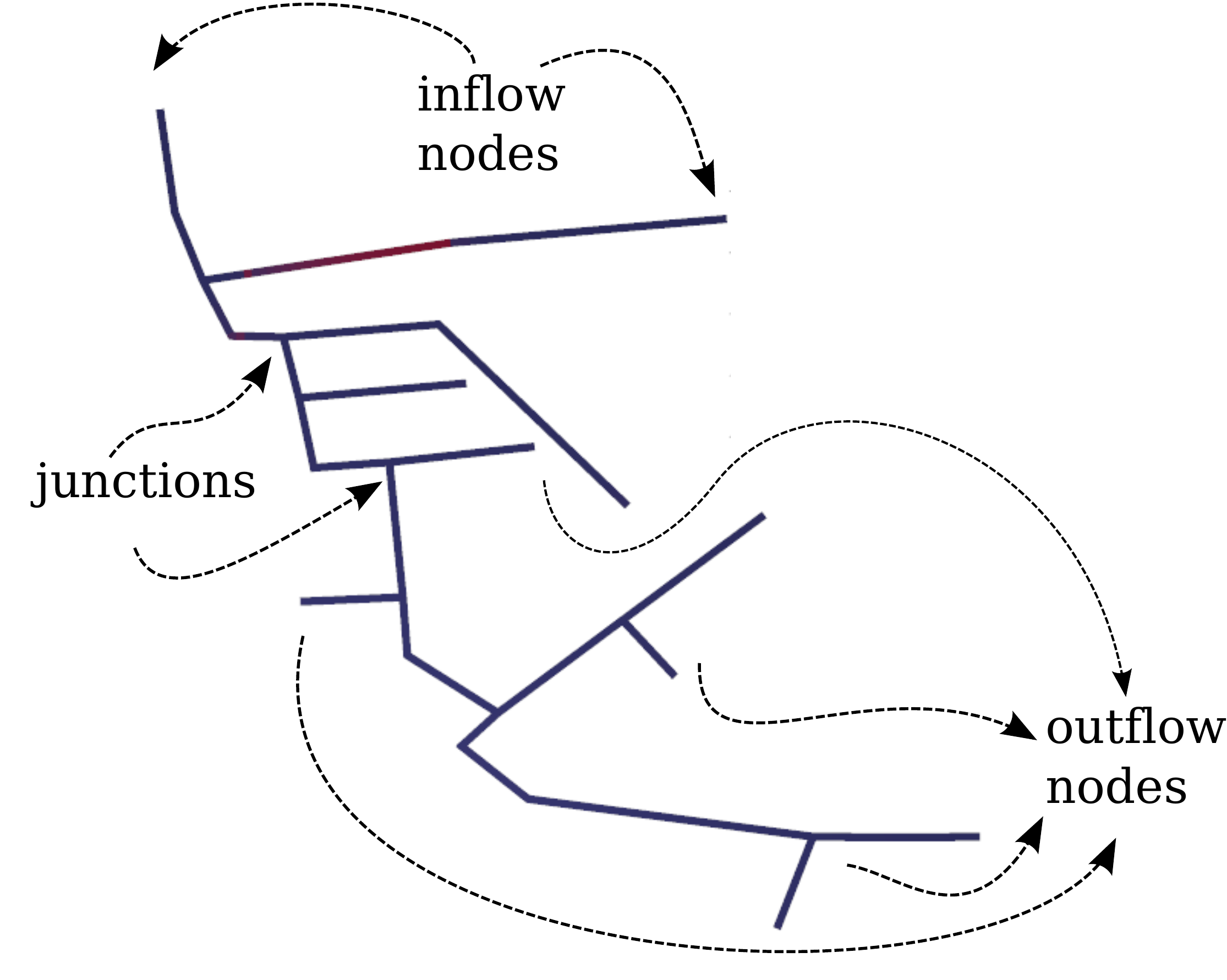} \\
      (a) & (b)
    \end{tabular}
  \end{center}
  \caption{(a) The Preissman slot model describes flow in a pipe of diameter $D$ by depth-averaging over this cross-section. Each filled cross-sectional area $A$ corresponds to a height $\xi$ and length $l(\xi)$. When $\xi$ exceeds a transition height $y_f$, water fills a narrow slot with width $T_s$ and contributes additional hydrostatic pressure. (b) An example pipe network. \label{fig:intro}}
\end{figure}

The modeled variables are the cross-sectional area $A$ and the discharge $Q$, which can be used to compute quantities of practical interest such as pressure head and velocity. Pressurization effects, though features of compressible flow, are accounted for by the Preissman slot cross section pictured in Figure \ref{fig:intro}(a). Above a transition height $y_f$ near the physical pipe crown, water fills the narrow slot, contributing hydrostatic pressure that may be interpreted as the pressure within the full pipe. The slot width $T_s$ is related to the effective pressure wave speed $a$ in the pressurized pipe via $a^2 = g A_f/T_s$, where $g$ is the acceleration due to gravity and $A_f$ is the cross-sectional area at the transition height $y_f$ (determined by the pipe diameter and $T_s$ ).  The pressure wave speed $a$ is, to first order, a function of pipe material only. In practice, the slot width $T_s$ is a parameter chosen to set the value of pressure wave speed $a$, which may range from 20--1250~m/s \cite{Vasconcelos2006} depending on practical context and numerical constraints. In each pipe, the governing equations for $A$ and $Q$ are the de St.~Venant equations for free-surface flow~\cite{Kerger2011,Leon2010},
\begin{equation}
\vq_t +(\vF(\vq))_x = \vec{S}, \quad 0< x<L,\quad 0<t<T,
\end{equation}
where $L$ is the length of the pipe, $T$ is the duration of the study period, and
\begin{equation}
\vq =  \bp A\\ Q \ep, \quad \vF = \bp Q\\ \frac{Q^2}{A} +gI(A)\ep,\quad \vec{S} = \bp 0\\S\ep,
\end{equation}
where $g$ is the acceleration due to gravity and the pressure contribution $I(A)$ is given by
\begin{equation}
I(A) = \int_0^{h(A)} (h(A) -\xi)l(\xi)d\xi,
\end{equation}
where $l(\xi)$ is the pipe width at height $\xi$. The quantity $\bar{p} = gI(A)/(A)$ is the average hydrostatic pressure over a cross-section. In the Preissman slot description, the pressure head $H$ is entirely captured by the hydrostatic term $\bar{p}$, and given by
\begin{equation}
H  \equiv \frac{\bar{p}}{\rho g} = \frac{I(A)}{A}.
\end{equation}
The friction term $S$ is
\begin{equation}
S = (S_0-S_f)gA,
\end{equation}
where $S_0$ is the slope of pipe bottom and $S_f$ empirically accounts for friction losses. In what follows we use the Manning equation
\begin{equation}
S_f =\frac{M_r^2 Q|Q|}{A^2 R_h(A)^{4/3}}
\end{equation}
where $R_h(A) = A/P_w$ is the hydraulic radius, which depends on the wetted perimeter $P_w$, which is a function of $A$ and $D$. The constant $M_r$ is the Manning roughness coefficient, which has an empirical value depending on pipe material. Other empirical formulae such as Hazen--Willams or Darcy--Weisbach may also be used. Initial conditions $\vq(x,0)$ are assumed to be known, and boundary conditions $\vq(0,t)$ and $\vq(L,t)$ are assigned based on the network connectivity and external inputs described in Section~\ref{sec:juncs}.

The Preissman slot approach has been used, for example, by Trajkovic et al.~\cite{Trajkovic1999}, who compared the model with experimental data; by Kerger et al.~\cite{Kerger2011}, who implemented an exact Riemann solver and introduced a ``negative slot'' modification to handle subatmospheric pressure; by Le\'on et al.,~\cite{Leon2009}, who allowed the pressure wave speed to vary slightly; and by Borsche and Klar~\cite{Borsche2014}, who used a hexagonal cross-section to simplify the area--height relationship. Other transient flow models for air and water in single pipes include a ``two-component pressure'' approach~\cite{Vasconcelos2006,Vasconcelos2011}; a single-equation
model with a modified pressure term~\cite{Bourdarias2007a,Bourdarias2008}; a two-component model~\cite{Leon2010}; and a three-phase model accounting for air, air--water mixture, and water~\cite{Kerger2012}.

Note that the de St.~Venant equations themselves make no assumptions about the channel cross section. The numerical methods in Section~\ref{sec:nums} may be rather easily adapted to other cross-sectional geometries, and in our implementation we also include the option to simulate flow in channels with uniform cross-sections. Our implementation could therefore be used to simulate other networks like rivers or irrigation canals, but such options have not yet been explored.

\section{Numerical Methods}
\label{sec:nums}
\subsection{Pipes}

\begin{figure}[ht]
  \begin{center}
\includegraphics[width =.8\textwidth]{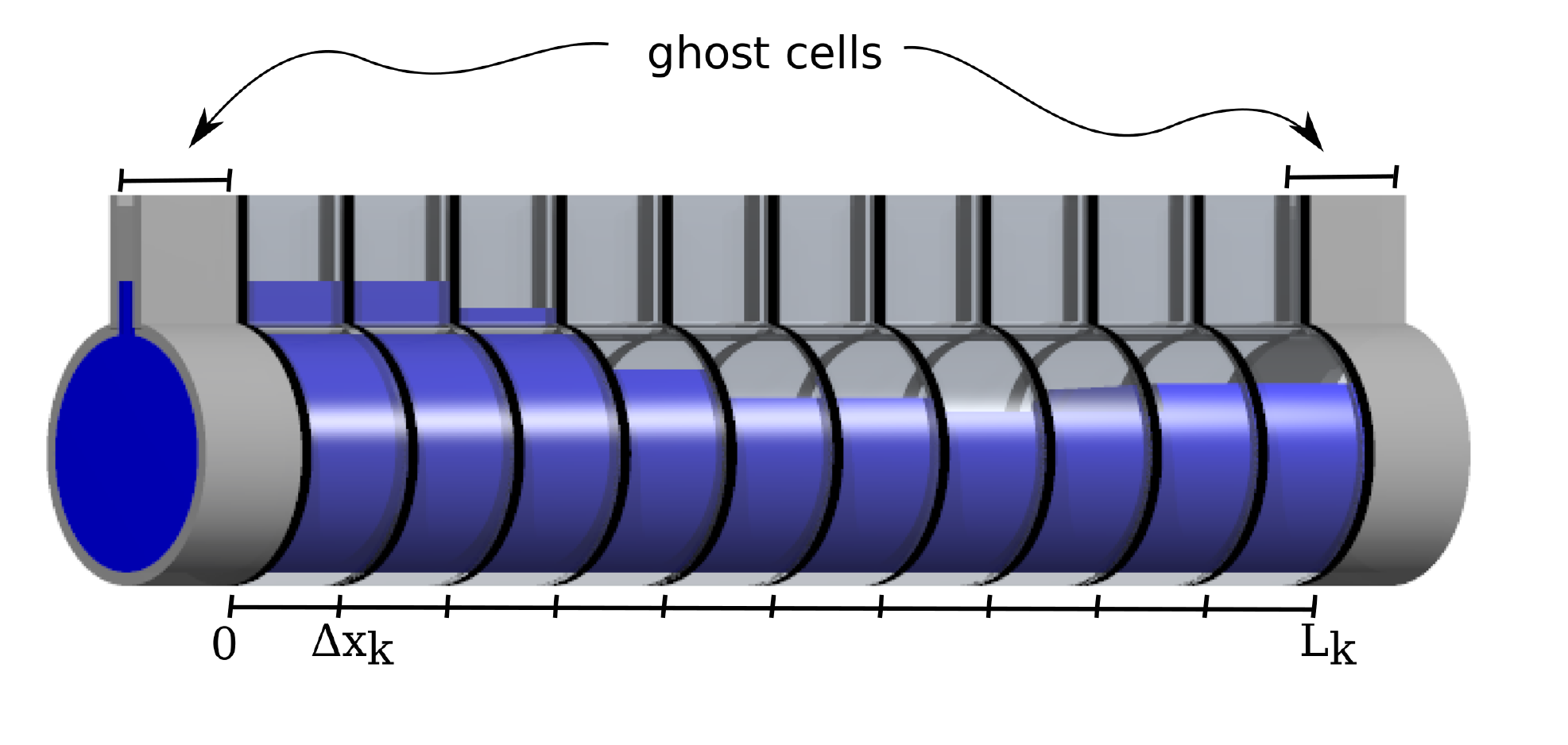}
\end{center}
\caption{Spatial grid layout for finite volume method in pipe $k$ of length $L_k$.}
\label{ps_grid}
\end{figure}

Each pipe $k$ is assigned a local coordinate $0\leq x_k \leq L_k$ and divided into $N_k$ cells of width $\Delta x_k$, where the cell centers are at \smash{$x_{k,j}=j+\frac{1}{2}\Delta x_k$} and the cell boundaries are at \smash{$x_{k,j\pm1/2}=x_{k,j}\pm\frac{1}{2}\Delta x_k $} for \smash{$j=0,\ldots, N_k-1$}, as shown in Figure \ref{ps_grid}. Each pipe is also padded with ghost cells at $x_{k,-1}$ and $x_{k,N_k}$ that are used to set the boundary conditions, described in Section \ref{sec:juncs}. A total of $M$ timesteps of size $\Delta t$ are taken.

 In what follows we drop the $k$ subscript and assume we are working within a single, specified pipe. Cell averages of area and discharge \smash{$(A^n_{j},Q^n_{j})\equiv \vq^n_{j}$} at the \smash{$n^\text{th}$} timestep are updated with an explicit third-order Runge--Kutta total variation diminishing (TVD) scheme~\cite{Gottlieb1998} that may be written in terms of first-order Euler update steps $E(\vq)$ as
\begin{equation}
  \label{eq:tvd}
  \hat{\vq} = \frac{3}{4} \vq^n+\frac{1}{4}E(E(\vq^n)),\qquad
  \vq^{n+1} = \frac{1}{3} \vq^{n} +\frac{2}{3}E(\hat{\vq}).
\end{equation}
Each Euler step $E(\vq)$ consists of updating the conservation law terms, the source terms, and the ghost cell values. For the interior cells, the update is
\begin{equation}
  E(\vq) = E_s(E_c(\vq))
\end{equation}
where the subscript $c$ denotes the conservation law update and the subscript $s$ denotes the source term update. First we treat the homogeneous conservation law term
\begin{equation}
  \vq_t +(\vF(\vq))_x = 0
\end{equation}
with a Godunov update
\begin{equation}
  E_c(\vq^n_j) = \vq^{n}_j-\frac{\Delta t}{\Delta x}\left(\vF_{j+\frac{1}{2}}-\vF_{j-\frac{1}{2}}\right).
\end{equation}
For the numerical flux function $\vF$, we use a Harten--Lax--van Leer (HLL) Riemann solver similar to that of Le\'on et al.~\cite{Leon2009}, which approximates the solution to the Riemann problem between a pair of cells (with left and right states $\vq_L$ and $\vq_R$ respectively) with a center state $\vq_*$ separated from the left and right states by shocks with speeds $s_L$ and $s_R$, respectively~\cite{Leveque2002}.

The expression for the center state flux $\vF_* = \vF(\vq_*)$ is found by applying the Rankine--Hugoniot condition across each shock. The Godunov update for this scheme is found by sampling this solution structure to obtain
\begin{equation}
  \vF_{j\pm\frac{1}{2}} =
  \begin{cases}
    \vF_L = \vF(\vq_L)& \text{if $s_L>0$,} \\
    \vF_*=\frac{ s_R\vF_L - s_L\vF_R+s_R s_L(\vq_R-\vq_L)}{s_R-s_L} & \text{if $s_L\leq 0 \leq s_R$,}\\
    \vF_R =\vF(\vq_R) & \text{if $s_R<0$,}
\end{cases}
\end{equation}
where
\begin{equation}
  (\vq_L, \vq_R) =
  \begin{cases}
    (\vq_j, \vq_{j+1})& \text{at $j+\frac{1}{2}$,} \\
    (\vq_{j-1}, \vq_j)& \text{at $j-\frac{1}{2}$.} \\
  \end{cases}
\end{equation}
We computed the shock speeds via $$s_L = u_L -\Omega_L, \quad s_R = u_R
+\Omega_R$$ where $u_j = A_j/Q_j$ is the velocity in cell
$j$ and
\begin{equation}
  \label{eq:omegaj}
  \Omega_j =
  \begin{cases}
    \sqrt{\frac{(gI(A_*)-gI(A_j))A_*}{A_j(A_*-A_j)}} & \text{if $A_*>A_j+\epsilon$,} \\
    c(A_*) & \text{if $A_*\leq A_j+\epsilon$,} \\
  \end{cases}
\end{equation}
where \smash{$c(A) = \sqrt{g A/l(A)}$} is the gravity wave
speed~\cite{Leon2009}. The center state is approximated by linearizing the
equations to obtain
\begin{equation}
A_* = \frac{A_L+A_R}{2}\left(1+\frac{u_L-u_R}{2\bar{c}}\right),
\end{equation}
where $\bar{c} = (u_L+u_R)/2$. Note that if $A_j<A_*$, the separating wave is
not in fact a shock, but a rarefaction, and the expression for $s_j$ gives the
speed of the head (or tail) of the appropriate rarefaction
wave~\cite{Leon2006}. A tolerance of $\epsilon =10^{-8}$ is incorporated into
\eqref{eq:omegaj} to account for the possibility that when $A_*$ and $A_j$ are
very close, small errors in the evaluation of $I(A)$ may cause the expression
underneath the radical to become negative. By Taylor expanding the expression
underneath the radical at $A_j$, one can verify that the two cases in
\eqref{eq:omegaj} connect continuously, and thus the incorporation of
$\epsilon$ has a minimal effect on the computation.
The shock speeds $s_L$ and $s_R$ may also be estimated by evaluating $\hat{u}$
and $c(\hat{A})$ for Roe averages $\hat{A}$ and $\hat{u}$ and taking minima and
maxima for left and right waves, respectively~\cite{Sanders2011}, but the
authors found the dynamics of interest in the present work were less robust
with this treatment.

Computing the HLL fluxes requires frequently evaluating the pressure term $I(A)$, the wave speed $c(A)$, and the integral $\phi(A)$, which are not analytic expressions of $A$ for the Preissman slot geometry. To avoid rootfinding at every cell at every time step, Chebyshev polynomials were used to accurately express these functions of $A$ and their inverses. Several of the functions are ill-suited to polynomial approximation due to fractional power singularities at either end of their domain, so we implemented an accurate interpolation by expanding in a series involving the relevant fractional power (Appendix \ref{app:cheby}).

The friction and slope source term updates, which only affect the second component, take the form
\begin{equation}
E_S(Q^n_j) = Q^n_j +\Delta t \, S\left(\vq^n_j+\frac{\Delta t}{2}S(\vq^n_j)\right).
\end{equation}

\begin{figure}
  \begin{center}
    \includegraphics[scale = 0.7]{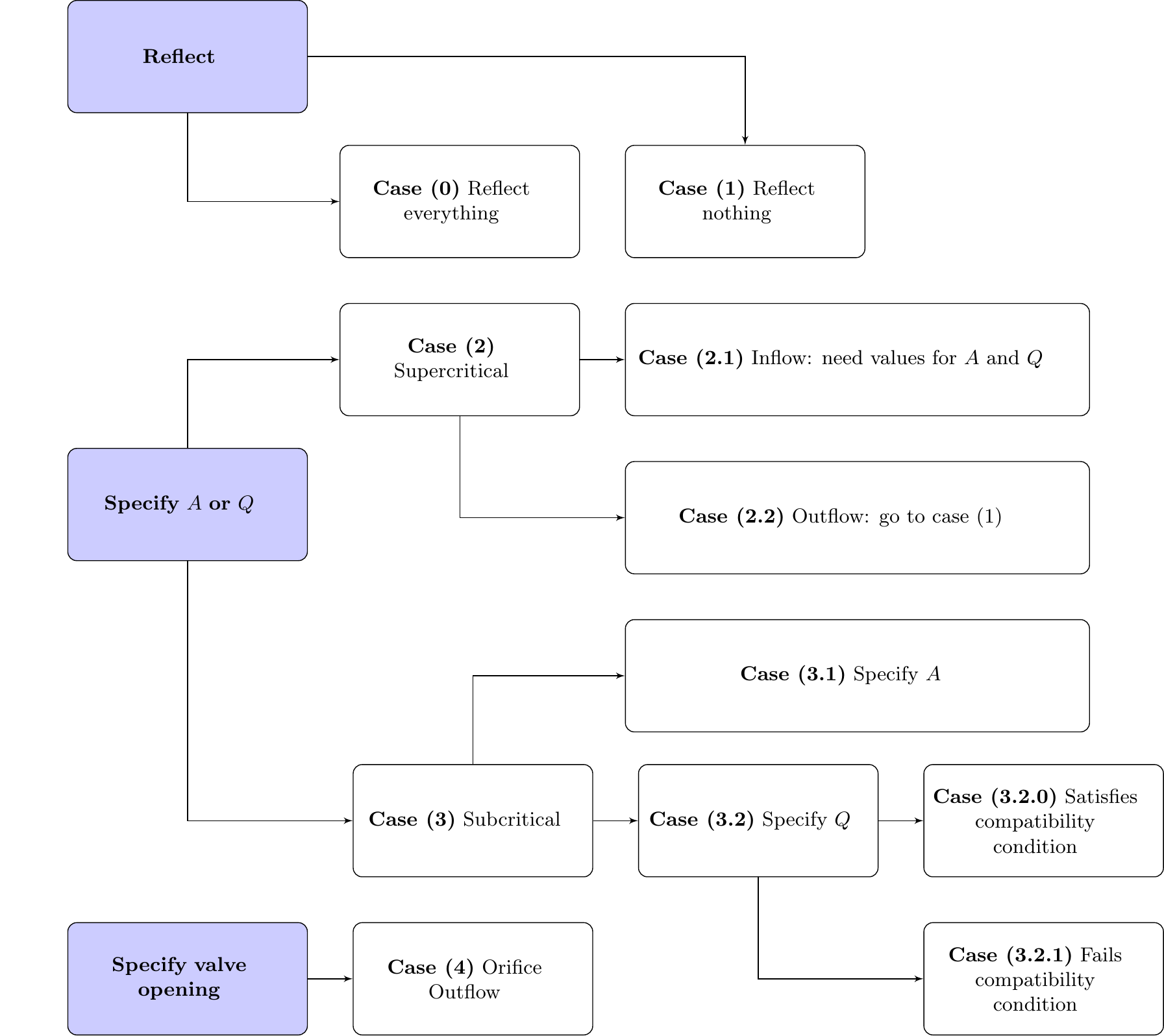}
  \end{center}
  \caption{Single pipe boundary cases. User may specify either reflection of all or no waves, a value for one of the dynamical variables $A$ or $Q$, or a valve opening.\label{fig:bcc}}
\end{figure}

\subsection{Junctions}
\label{sec:juncs}
Each pipe domain is padded with ghost cells, which serve to compute fluxes to update boundary cells in each pipe's computational domain. In what follows, we use the notation $(A^k_\te, Q^k_\te)$ to denote the ghost cell values of $A$ and $Q$ at the relevant end of pipe $k$ and denote the values of the last cell in the computational domain by $(A^k_\ti, Q^k_\ti)$. The algorithm uses the network layout to determine what update routine is used on the ghost cell values $(A^k_\te, Q^k_\te)$. The update routines fall into three categories: external boundaries, two-pipe junctions, and three-pipe junctions. Note that the user must specify additional information for the external boundary routine. For other boundary treatments in networks, see~\cite{Sanders2011},~\cite{Capart1999}, and~\cite{Leon2010a}.

\subsubsection{External boundary routine}
When one end of a pipe is connected to the outside world, the user may specify a variety of cases, summarized in Figure \ref{fig:bcc}, to describe the external conditions. In each case, the user specification of boundary condition type allows the solver to update the ghost cell values. For example, in the network shown in Figure \ref{fig:intro}(b), a time series of $Q$ or $A$ would be specified at each ``inflow'' node (cases (2) and (3)). At the labeled ``outflow'' nodes, the user may choose between several possible descriptions of how a valve allows water to exit the end of a pipe. These boundaries may be considered either as either orifice outflow (case (4)) or unimpeded outflow where no waves are reflected (case (1)). A closed valve may be simulated as a reflective boundary (case (0)).

As only one pipe is involved, in what follows we drop the $k$ superscript on the ghost and internal cell values. For cases (0) and (1), the ghost cell values $(A_\te, Q_\te)$ are updated via
\begin{equation}
(A_\te,Q_\te)=
  \begin{cases}
     (A_\ti,-Q_\ti) &\text{case (0):  reflect everything,}\\
    (A_\ti,Q_\ti)&\text{case (1): reflect nothing.}
  \end{cases}
\end{equation}
Case (0) reflects all waves and case (1) reflects none \cite{Leveque2002}. Physically, case (0) corresponds to a dead end and case (1) corresponds to an opening with unimpeded outflow.
 
Cases (2) and (3) arise when the user specifies a time series for either $A_\te(x,t)$ or $Q_\te(x,t)$ at the external boundary. During each Euler update step, the solver first determines whether the interior flow is super- or subcritical, depending on whether the Froude number $F_r= u/c(A)$ is greater than or less than unity, respectively.

In the supercritical case (2), if $Q_\te<0$ at $x=0$ or $Q_\te>0$ at $x=L$, then outflow case (2.1) applies. Information from the boundary cannot propagate inside the domain under these conditions, and thus case (2.2) is evaluated in the same manner as the extrapolation case (1), where all outgoing waves continue untrammeled. For supercritical inflow, case (2.1) applies. If $Q_\te$ is specified, the undetermined ghost cell value of $A_\te$ is set to the initial inflow cross-sectional area; otherwise the Froude number in the ghost cell is set equal to the Froude number just inside the domain.

For subcritical flow, information may propagate in both directions, and the solver determines a value for the unspecified component $A_\te$ or $Q_\te$. The approach in the current work is to attempt to follow an outgoing characteristic and use the value of the Riemann invariant along this characteristic to solve for the unknown external value. Sanders and Katopodes~\cite{Sanders2000} used an exact version of this for networks of channels with uniform cross-section (in this case the Riemann invariants are simple). The same idea was implemented iteratively for a circular geometry by Le\'on et al.~\cite{Leon2006} and Vasconcelos et al.~\cite{Vasconcelos2006}. However, care must be taken to ensure that the characteristic assumption is valid. Thus our algorithm for case (3.1) works as follows:
\begin{remunerate}
  \item Calculate the Riemann invariant for the last cell in computational domain
    according to $R^{\pm} = Q_\ti/A_\ti \pm \phi(A_\ti)$, where $(A_\ti,
    Q_\ti)$ are values in the last cell, and $\phi$ = $\int (c/A) dA$ depends
    on geometry. \smash{$\phi(A) = 2\sqrt{gA/w}$} for uniform cross-sections of
    width $w$, and has no analytic expression for circular cross sections (see
    Appendix \ref{app:cheby} for Chebyshev representation).
  \item[2a.] If $A_\te$ is specified, set $$Q_\te = A_\te
    \left(Q_\ti/A_\ti\pm(\phi(A_\ti)- \phi(A_\te))\right),$$ where the solution
    is physically valid since $Q$ is allowed to be positive or negative. Even
    though a solution will always be found, it may violate the subcritical
    condition.
  \item[2b.] If $Q_\te$ is specified, check for compatibility as described
    below. If incompatible, go to case (3.2). Otherwise, rootfind to solve for
    $A_\te$ satisfying
    \[
    Q_\ti/A_\ti \pm \phi(A_\ti) = Q_\te/A_\te \pm \phi(A_\te).
    \]
  \item[3.] Update ghost cell values to $(A_\te, Q_\te)$.
\end{remunerate}
Rootfinding in step (2b) above is by no means guaranteed to work; indeed, a positive solution $A_\te$ only exists for certain combinations of the parameters. Physically, this means that not all values of $Q_\te$ may be continuously connected to the interior state, and choosing certain $Q_\te$ forces a shock between the ghost cell and the last computational cell. For the boundary at $x=0$, for a given value of $Q_\te<0$ there is a maximum allowed value of $c_{-} =  Q_\ti/A_\ti - \phi(A_\ti)$. Similarly, for the boundary at $x = L$, for a given $Q_\te>0$, there is a minimum allowed value of $c_{+} =  Q_\ti/A_\ti + \phi(A_\ti)$.
For the uniform cross-section case, one can find analytic formulas for these maximum/minimum allowed values:
\begin{equation}
c^{*}_{-} =   -\left(\frac{g}{l}\right)^{1/3}|Q_\te|^{1/3}, \qquad
c^{*}_{+} = 3\left(\frac{g}{l}\right)^{1/3}(Q_\te)^{1/3}.
\end{equation}
For the Preissman slot, to find the values of $c^{*}_{\pm}$ we estimate the critical point of the function
$g(t) = Q_\te/t\pm \phi(t)$
rather than rootfind for the exact value. We define \smash{$c^*_{\pm} = Q_\te/\hat{x} \pm \phi(\hat{x})$}, where \smash{$\hat{x}^3 = \frac{D}{g}Q_\te^2 $} (for pipe diameter  $D$), which corresponds to approximating the gravity wave speed by \smash{$\sqrt{gh(A)}$}. The compatibility condition is then
\begin{equation}
\text{if }\left\{\begin{array}{c} c_{-} >  c^{*}_{-}\\c_{+} <  c^{*}_{+}\end{array}\right\}  \text{ at } \left\{\begin{array}{c} x =0\\x=L\end{array} \right\}
\text{ then } Q_\te \text{ is not compatible.}
\label{eq:qcomp}
\end{equation}

Should $Q_\te$ fail the compatibility condition, case (3.2) applies, and the solver sets $A_\te=A_\ti$. This choice causes the HLL center state estimate $A^*< A_\ti= A_\te$, consistent with states separated by shocks in the eyes of the approximate Riemann solver update.

Lastly, if a valve or gate opening is specified, case (4) applies. Bernoulli's equation applied to flow through an orifice gives
\begin{equation}
Q_\te = C_d A \sqrt{2g(h_\ti-C_c h_\te)},
\end{equation}
where $h_\ti=h(A_\ti)$, $h_\te$ is a height between 0 and $D$ that indicates the valve opening, $g$ is the acceleration due to gravity, and $A = A(h_\te)$ is the area of the outflow orifice. The discharge coefficient $C_d=0.78$\, and the contraction coefficient $C_c= 0.83$, are empirical constants from experiment~\cite{Trajkovic1999}. In this case, $A_\te = A_\ti$.

\begin{figure}
  \begin{center}
  \includegraphics[width = 0.95\textwidth]{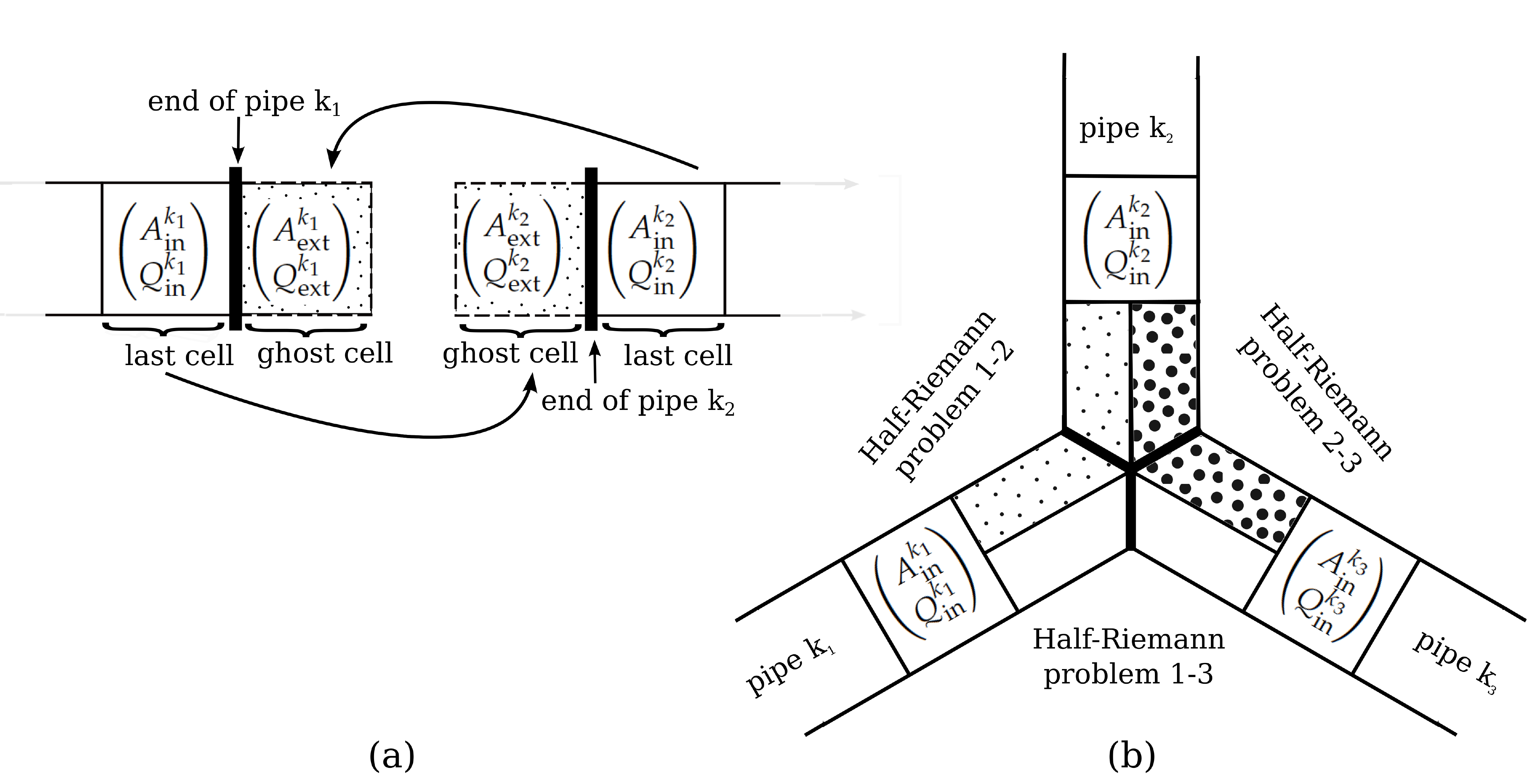}
  \end{center}
  \vspace{-1mm}
  \caption{Junction routine schematics for (a) two pipes, and (b) three pipes. \label{fig:junc}}
\end{figure}

\subsubsection{Two-pipe junction routine}
In the absence of valves, we apply mass conservation and assert that water height is constant across a two-pipe junction. Hence, the ghost cell in pipe $k_1$ is updated by translating water height from the pipe $k_2$ to the local geometry in pipe $k_1$ and copying the discharge from pipe $k_2$. That is, for example, pipe $k_1$ gets ghost cell values $(A^{k_1}_\te, Q^{k_1}_\te)$ such that $h_{k_1}(A^{k_1}_\te) = h_{k_2}(A^{k_2}_\ti)$ and $Q^{k_1}_\te=Q^{k_2}_\ti$. This procedure is shown in Figure \ref{fig:junc}(a).

\subsubsection{Three-pipe junction routine}
Triple junctions are divided into three pairs of two-pipe subproblems, which are solved to find fluxes. Each two-pipe subproblem contributes half the flux to an incoming pipe, as shown in Figure \ref{fig:junc}(b). For example, the fluxes at the end of pipe $k_1$ are the sum of half the flux from the two-pipe junction routine between $(A^{k_1}_\ti, Q^{k_1}_\ti)$ and $(A^{k_2}_\ti, Q^{k_2}_\ti)$ and half the flux from the two-pipe junction routine between $(A^{k_1}_\ti, Q^{k_1}_\ti)$ and $(A^{k_3}_\ti, Q^{k_3}_\ti)$.  If the coordinate systems do not point in the same direction (e.g. both pipes have $x=0$ at the triple junction), then one of the discharge terms must have a relative minus sign before the two-pipe junction routine is applied. The flux assignment does not otherwise depend on geometry. 

The authors believe this formulation is a simple way to couple the one-dimensional problems with minimal computational effort, since it recycles the two-junction solver and introduces no new data structures or solution routines. This routine may be improved by accounting for geometric effects and energy losses (often referred to as ``minor losses'' in pipe flow parlance) due to the specific geometry of the junction. Other treatments of these types of boundaries include Le\'on et al.~\cite{Leon2010a}, in which the triple junction has finite area and a dropshaft, and $n$-pipe junction implementations~\cite{Colombo2008,Borsche2014}.

\subsection{Model implementation}
The computational model (provided in supplementary materials) is written in
C++, and the different simulations are initialized through two text files: an
EPANET-compatible file describing network layout, and a configuration file
specifying additional simulation parameters. In addition, a Cython wrapper is
available, allowing simulations to be launched from iPython notebooks. The
simulation running time analyses that are reported in the following sections
were performed on a Mac Pro (Mid 2010) with dual 2.4~GHz quad-core Intel Xeon
processors, using a single thread unless stated otherwise. The authors found
that parallelizing the network using the OpenMP library (so that all pipes are
solved simultaneously) offered no advantage, because doing single-timestep
Riemann solver updates along each pipe is too fast to make it worthwhile to
spawn separate threads.

\section{Model validation and results}

\subsection{Experiments of Trajkovic et al.~(1999)}
Experiments were performed in a single pipe of flow transitioning from open-channel to pressurized and back~\cite{Trajkovic1999}. The experimenters used the Preismann slot to model their results, and their experiments have been revisited in several other studies~\cite{Leon2009,Vasconcelos2007}. The experiments were performed in a single pipe of length $L=10$~m and diameter $D = 0.1$~m, set at a slope of 2.7\% as reported in \cite{Trajkovic1999}. The Manning roughness coefficient was estimated as 0.002 after examining modeling results for a small range of roughness coefficient values. We consider the ``type A'' experiments where the initial condition was supercritical unpressurized flow with $h=0.014\text{~m}$ and $Q=0.0013\text{~m/s}$ throughout the pipe. At time $t=0\text{~s}$ a gate at $x=9.6\text{~m}$ is closed, causing the pipe to pressurize. At time $t=30\text{~s}$, the gate is partially reopened to a height $e_0$, which allows the flow to de-pressurize if $e_0$ is sufficiently large. This experiment was modeled with $\Delta x = 0.05\text{~m}$ and a Courant number of 0.6. Pressure traces at two sensors, P5 and P7, located at $0.5\text{~m}$ and $2.5\text{~m}$ upstream of the gate, respectively, are plotted in Figure \ref{fig:traj} for three values of $e_0$ are shown below. Comparison with experimental data shows that the model arrival times and  pressure traces are in good agreement. 
\begin{figure}
  \begin{center}
    \begin{tabular}{c}
      \includegraphics[width = .94\textwidth]{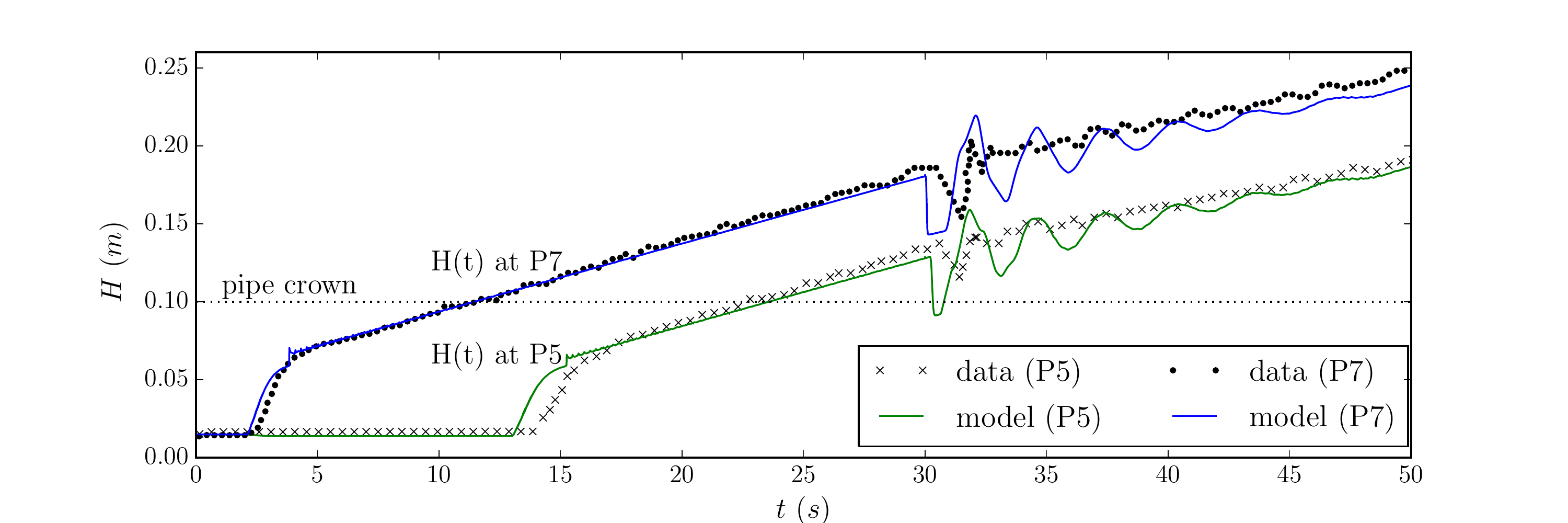}\\
      \includegraphics[width = .94\textwidth]{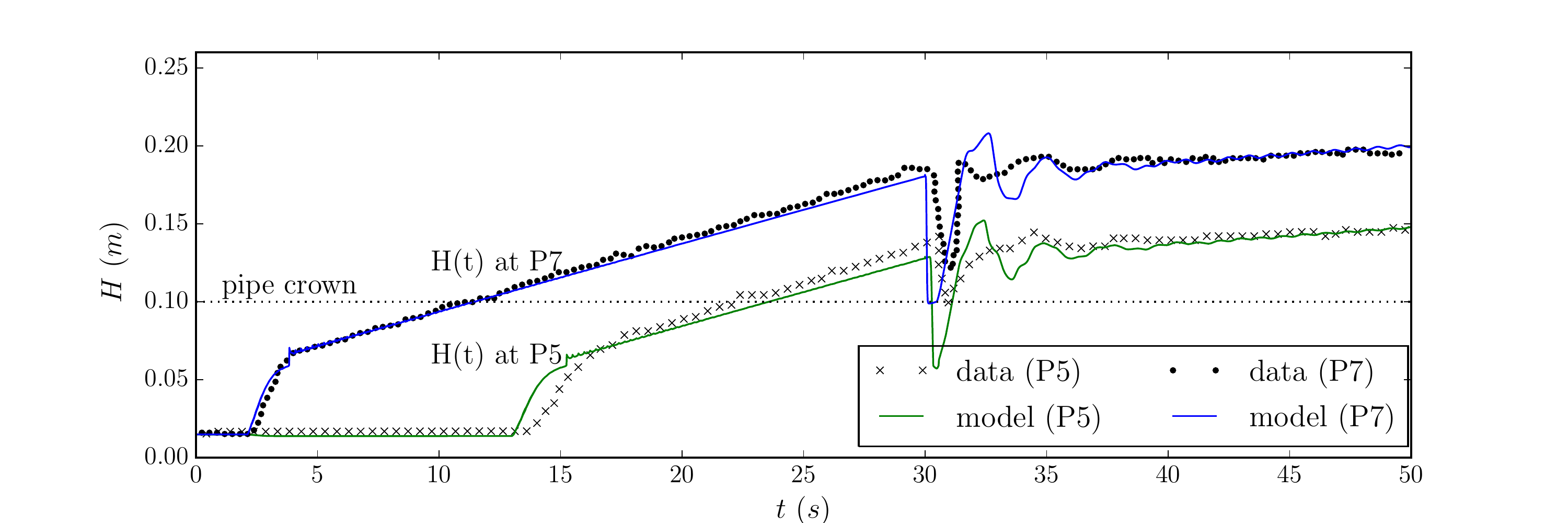} 
    \end{tabular}
  \end{center}
  \vspace{-1mm}
  \caption{Model (thin lines) and experimental~\cite{Trajkovic1999} (points) pressure head at locations P7 and P5, for gate openings $e_0=~0.008$~m (top) and $e_0=0.015$~m (bottom).\label{fig:traj}}
\end{figure}

\subsection{Water Hammer}
\begin{figure}
  \begin{center}
    \includegraphics[width = .94\textwidth]{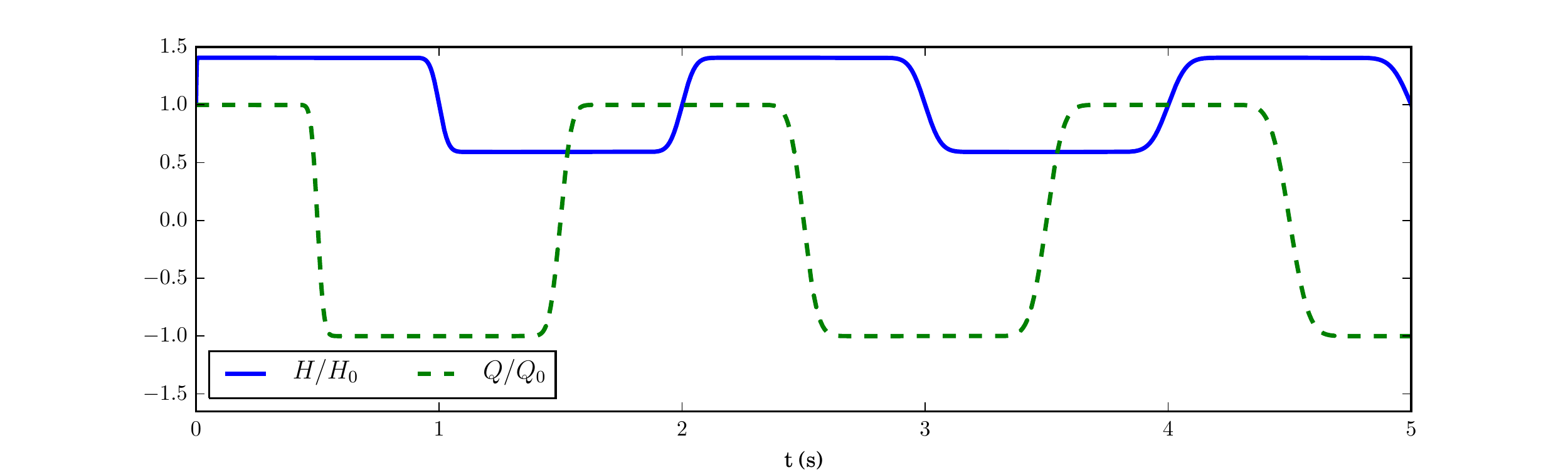}
  \end{center}
  \vspace{-1mm}
  \caption{Water hammer example. Time series for $H(x=L,t)/H_0$ and
  $Q(x=0,t)/Q_0$.\label{fig:hammer}}
\end{figure}

In this textbook example, similar to a case presented by Vasconcelos et al.~\cite{Vasconcelos2007}, a sudden valve closure gives rise to a so-called water hammer phenomenon in a pressurized pipe, whereby a shock wave of high pressure travels rapidly up and down the pipe. The setup for this problem consists of a horizontal frictionless pipe of length $L= 600\text{~m}$ and diameter $D=0.5\text{~m}$. Initially, the pipe has pressure head $H_0=150\text{~m}$, and a discharge $Q_0 = 0.1 \text{~m}^3\text{/s}$. The upstream pressure head at $x=0\text{~m}$ is held steady at 150~m, and a reflective boundary is applied at $x=L$. The pressure wave speed set by the Preissman slot is 1200~m/s. Results in Figure \ref{fig:hammer} show the time evolution of the nondimensionalized pressure head $H(x=L, t)/H_0$ at the valve and the discharge $Q(x=0,t)/Q_0$ at the upstream end.
The modeled results agree with the classical water hammer equations~\cite{Wylie1993}, which assert that for a pressure wave propagating in pressurized pipe, the relationship between the change in pressure head $\Delta H$ and the change in velocity $\Delta V= \Delta Q/A_f$ is
$$\frac{\Delta H}{\Delta V} = \frac{a}{g},$$
where $A_f$ is the cross-sectional area of the full pipe (corresponding to transition height $y_f$), $g$ is the acceleration due to gravity and $a$ is the pressure wave speed. With 600 grid cells and a Courant number of 0.6, we computed \smash{$|\frac{\Delta H}{\Delta V} -\frac{a}{g}| =0.00002$}.

\subsection{Grid refinement}
\begin{figure}
  \begin{center}
    \includegraphics[width = .8\textwidth]{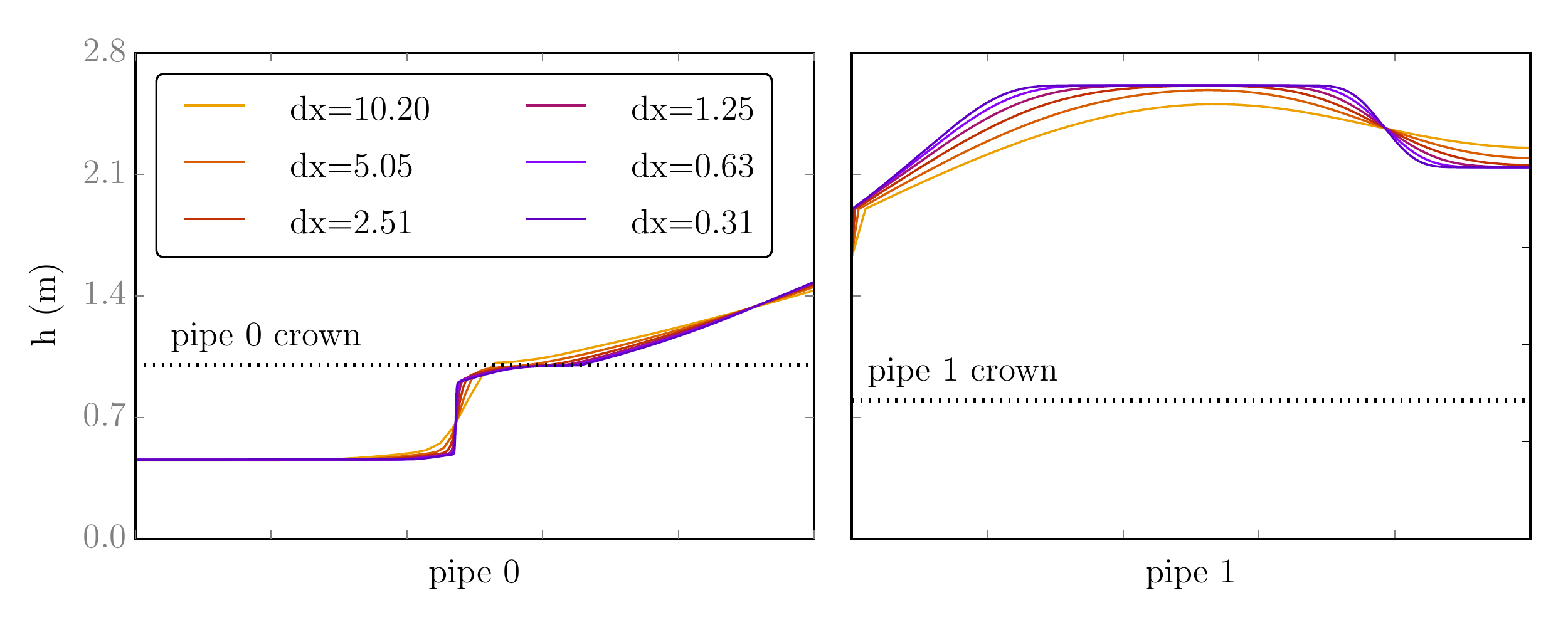}
  \end{center}
      \vspace{-2mm} 
  \caption{Grid convergence test,  water height $h(x,t)$ at $t=100$~s. \label{fig:conv2a}}
\end{figure}

We consider two horizontal, frictionless pipes connected in serial, each of length 500~m. The left pipe (denoted pipe 0) has diameter 1~m, and the right pipe (denoted pipe 1) has diameter 0.8~m, allowing for examination of the effect of a constriction upon the simulated flow. Starting with initial conditions $Q(x,t=0) = 1\text{~m}^3\text{/s}$ and $h(x,t=0)=0.75\text{~m}$ in both pipes, we run the simulation until $t=100\text{~s}$. The boundary conditions are specified inflow $Q(x=0,t)=0.1\text{~m}^3\text{/s}$ for pipe 0 and and reflection at $x=500\text{~m}$ for pipe 1. Pressurization occurs in part of the system, demonstrating the Preissman slot in action. Note that these examples used a pressure wave speed of 12~m/s corresponding to a relatively wide slot. Running convergence studies with a physical wave speed on the order of 1000~m/s would have given the same behavior but requires a much smaller timestep. We run the simulation with a range of values of $\Delta x$ to obtain the results pictured in Figure \ref{fig:conv2a}, which shows final pressure head $h(x,t=100)$ in both pipes. Table \ref{tab:conv1} shows  the error at time $T$, defined as $||e||_1= \sum_{j,k}|h^k_j-h^k(x_j,T)|\Delta x$, where $h^k(x_j,T)$ is the solution in pipe $k$ on the finest grid, evaluated at point $x_j$. As expected, we observe first-order convergence, due to the presence of shocks. The table also lists the computation times of the test,  The computation times scale quadratically with $N$ since the number of timepoints $M$ must also scale with $N$ to satisfy the CFL condition.

\subsection{Realistic network}
\begin{figure}
  \begin{center}
    \footnotesize
\includegraphics[width =.87\textwidth]{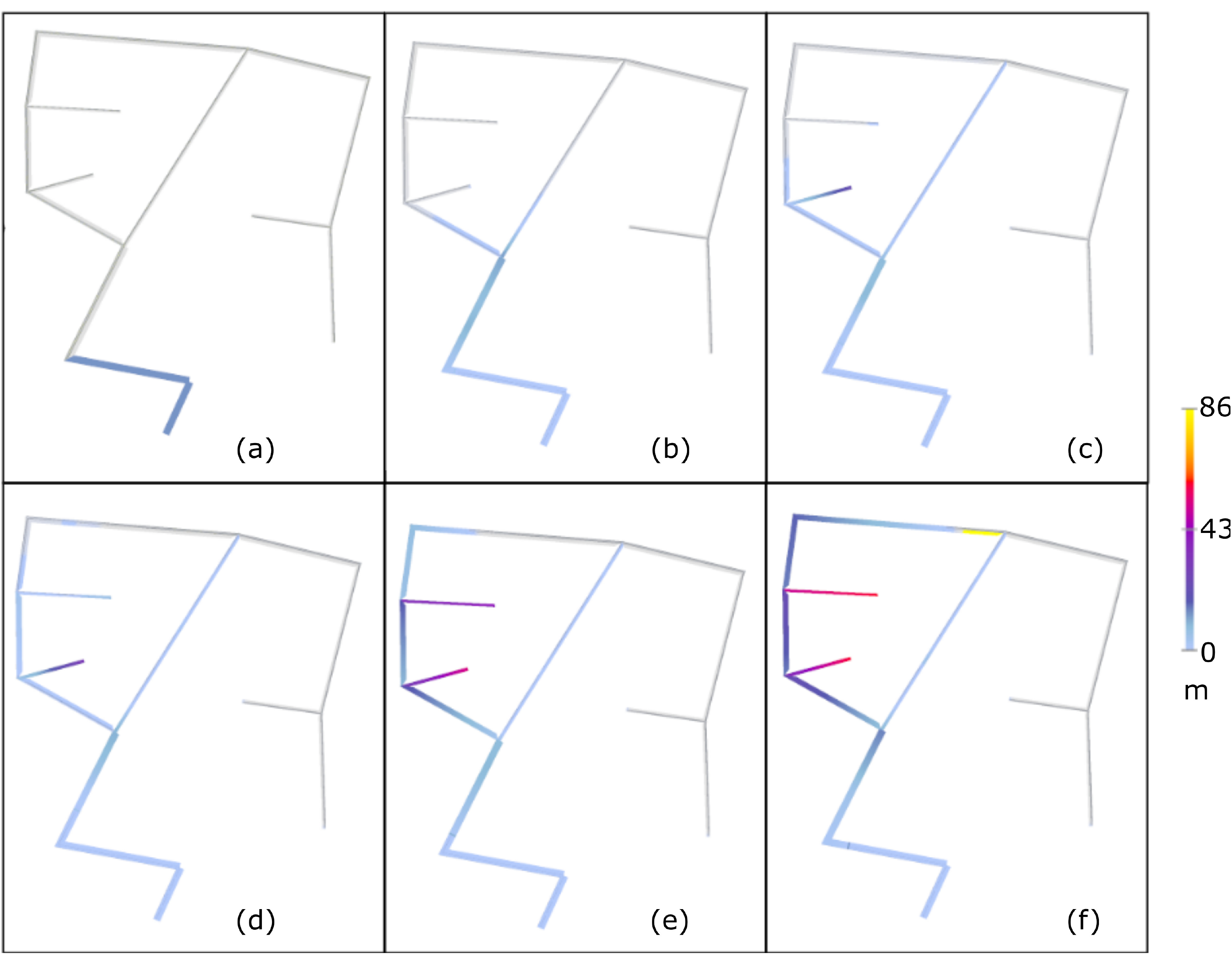}
  \end{center}
  \vspace{-1mm}
  \caption{Pressure head (m) in example (4.4). Grey pipes are empty. Times shown are (a) $t=0\text{~s}$  (b) $t=240\text{~s}$, (c) $t=480\text{~s}$, (d) $t=720\text{~s}$, (e) $t=960\text{~s}$, (f) $t=1200\text{~s}$. \label{fig:7sep}}
\end{figure}

We next consider, as an illustration of the model's ability to handle a realistic scenario, a larger network layout based on a network experiencing intermittent supply in a suburb of Panama City. The network contains fourteen pipes, ranging 300--1100~m in length and 0.4--1.0~m in diameter. Slope terms are non-zero in all pipes and Manning coefficient is estimated at $0.015$ based on pipe material. We simulate filling over a period of 1120 seconds, starting with initial conditions pictured in Figure \ref{fig:7sep}(a), to obtain the pressures in Figure \ref{fig:7sep}(b)--(f). The simulated pressure values fall within in a realistic range, as do the filling times of the pipes that pressurize. The fact that not all pipes pressurize during the simulation period is also commensurate with observations.

\begin{table}
  \caption{Convergence results, solution computed at time $t=100$ s. \label{tab:conv1}}
      \vspace{-2mm} 
  \begin{center}
    \footnotesize
    \begin{tabular}{|l|l|l|l|}
      \hline
      $N$ & $M$ & Wall clock time (s)& $||e||_1$ \\
      \hline
 50 &  200 & 0.10   &        0.18251     \\
100 &  400 & 0.40   &        0.10791     \\
200 &  800 & 1.54   &        0.05608     \\
400 & 1600 & 5.84   &        0.02188     \\
800 & 3200 & 23.80   &        0.01380     \\
      \hline
    \end{tabular}
  \end{center}
\end{table}

\begin{figure}
  \begin{center}
    \includegraphics[trim={0.2 0 0.2 4cm},clip,width = .75\textwidth]{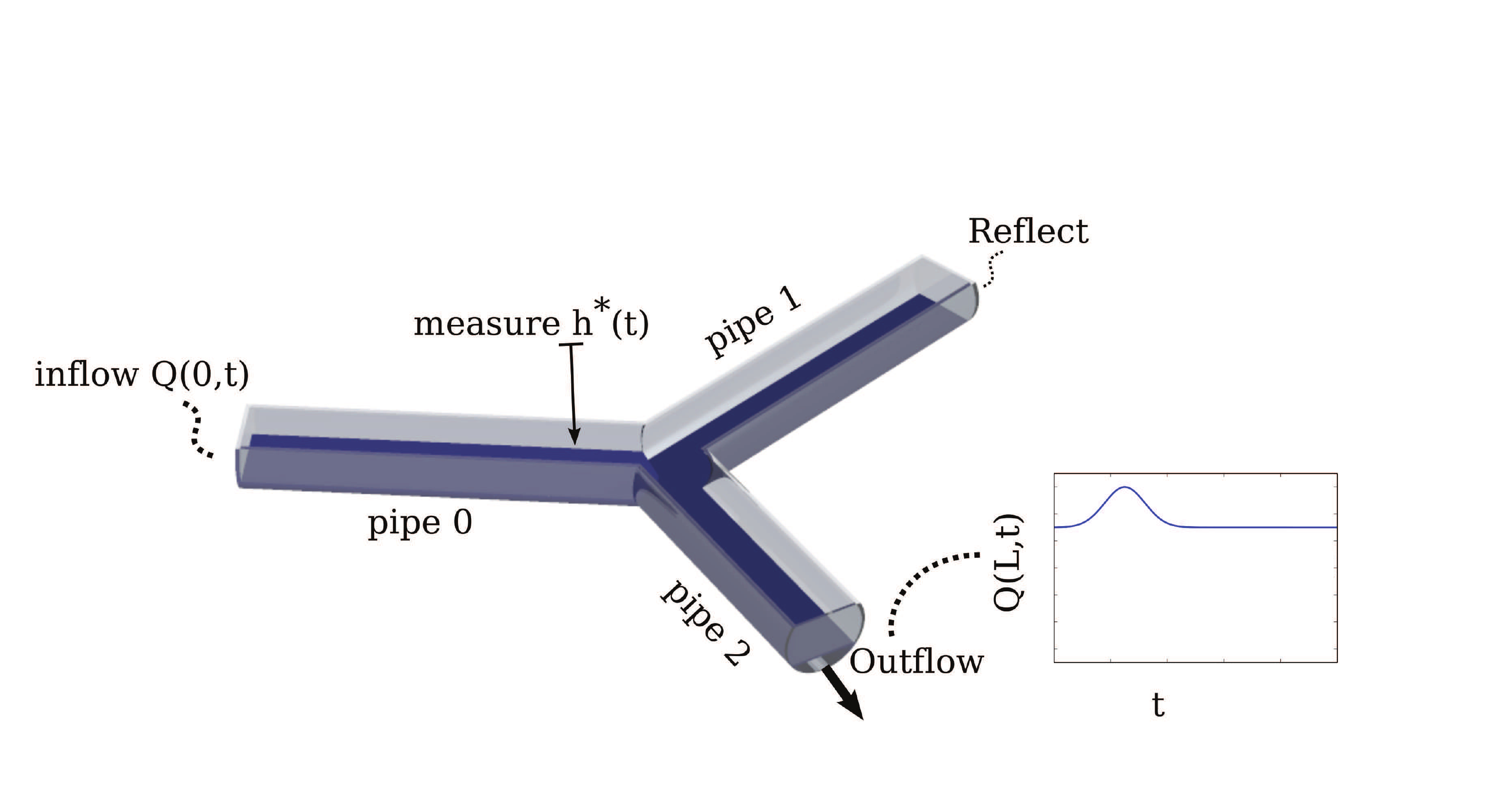}
  \end{center}
  \vspace{-8mm}
  \caption{Illustration of boundary recovery example. Inflow $Q(0,t)$ to pipe 0 is prescribed on the left. Measurements of $h(x^*,t)=h^*(t)$ in pipe 0 are used to find the unknown pipe 2 outflow $Q(L,t)$. \label{fig:diag4}}
\end{figure}

\section{Optimization}
\label{sec:optimize}
We now extend the computational model to consider several optimization problems that are inspired by realistic scenarios. We use the Levenberg--Marquardt (LM) algorithm, a trust-region method, to determine an unknown quantity that minimizes an objective function. When the unknown quantity is a time series, as in Sections \ref{sec:opti1} and \ref{sec:opti2}, the degrees of freedom describe either a Hermite spline or a truncated Fourier series. The LM algorithm requires computing a Jacobian, and this was calculated using finite differences, since the nonlinearly coupled boundary conditions made a variational formulation impractical. The Jacobian columns are computed in parallel using the OpenMP library, which provided considerable time savings over a serial implementation for the problems considered.

\subsection{Recover unknown boundary data}
\label{sec:opti1}
\begin{figure}
  \begin{center}
    \footnotesize
    \begin{tabular}{cc}
      \includegraphics[width=0.465\textwidth]{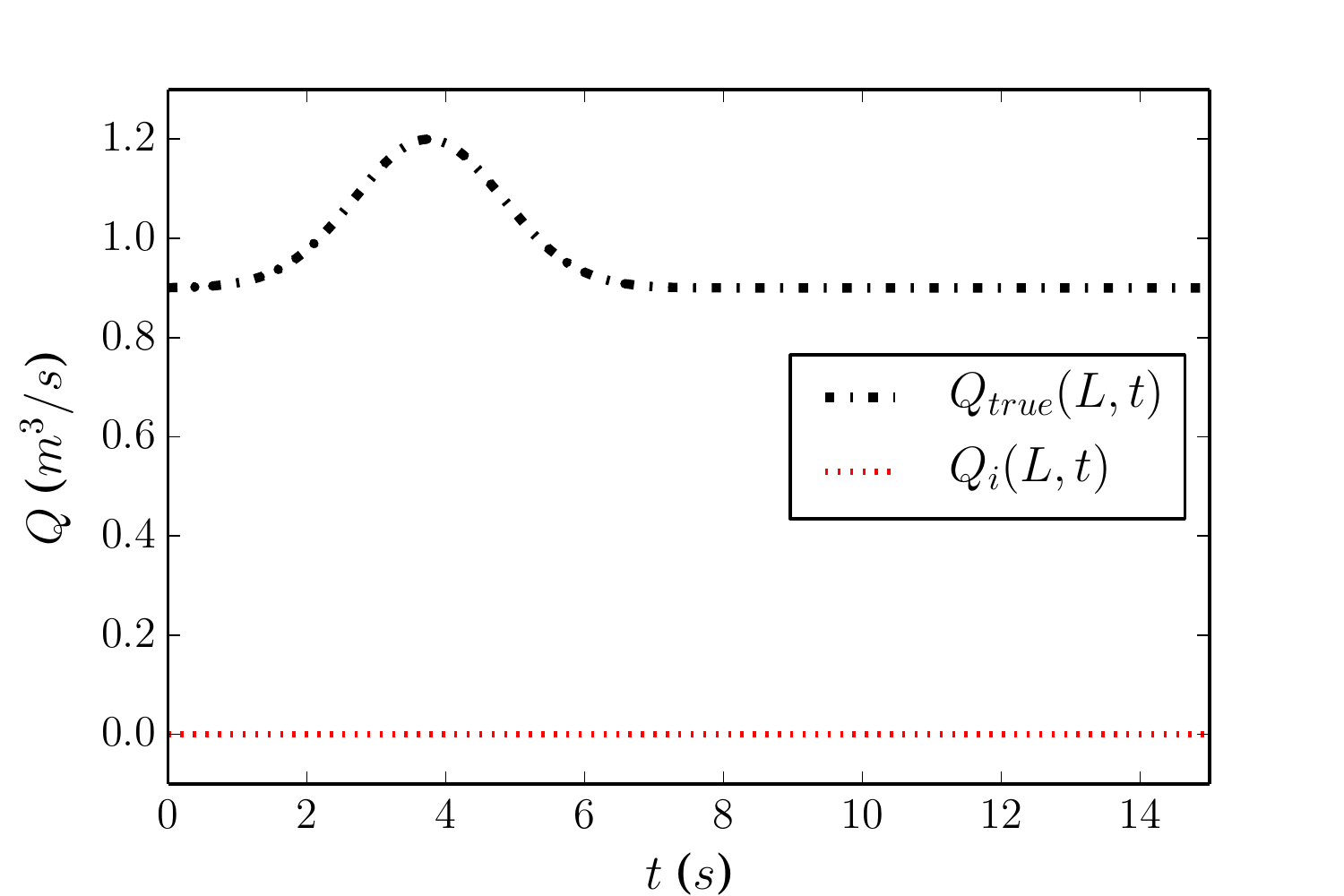} & 
      \includegraphics[width=0.465\textwidth]{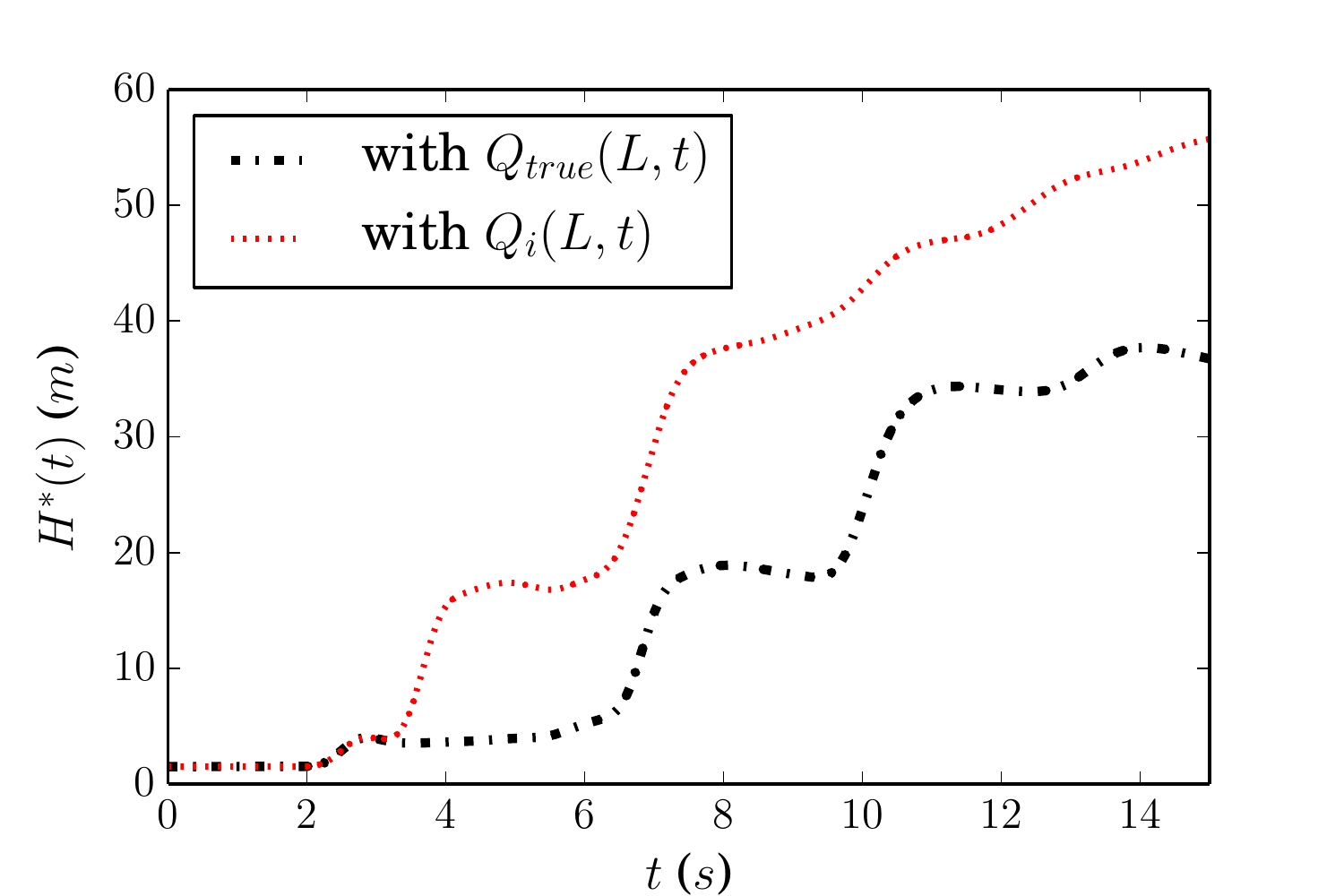} \\
      (a) & (b)
    \end{tabular}
  \end{center}
      \vspace{-2mm} 
  \caption{(a) Pipe 2 outflow times series $Q_\text{true}$ used to generate
  data. (b) Sensor measurement $H^*(t)$ when pipe 2 outflow is given by
  $Q_\text{true}$ (black dashed line) vs. $Q_i$ (red dash--dot
  line).\label{fig:whargarbl}}
\end{figure}

The uncertainty of intermittent water supply motivates our first example, in which we are interested in recovering unknown boundary information based on our knowledge of pressure somewhere in the network. Below, we provide proof-of-concept for the modeling/optimization framework by solving a problem where we know the correct answer. Figure \ref{fig:diag4} shows a diagram of the example network considered. The pipe diameters are all 1~m, and the lengths are 500~m, 500~m, and 125~m for pipes 0, 1, and 2, respectively, $\Delta x =2.5 \text{~m}$ for all pipes, and the number of time steps is $M = 700$. Inflow to pipe 0 is a constant $1\text{~m}^3\text{/s}$. Pipe 1 has a reflective boundary at the downstream end, and the outflow time series of pipe 2 is not known. We generate a test data set with the outflow time series $Q_\text{true}(L,t)$ as depicted in Figure \ref{fig:whargarbl}(a). The slot width is 0.00053~m, corresponding to a pressure wave speed of 120~m/s. 	
where $H(x^*,t_i)$ is the solution at the sensor at time step $i$. The degrees of freedom are Hermite spline coefficients or Fourier modes for the boundary value time series $Q(L,t)$ on the outflow end of pipe 2.

\begin{figure}
  \begin{center}
    \footnotesize
    \begin{tabular}{cc}
      \includegraphics[width=0.465\textwidth]{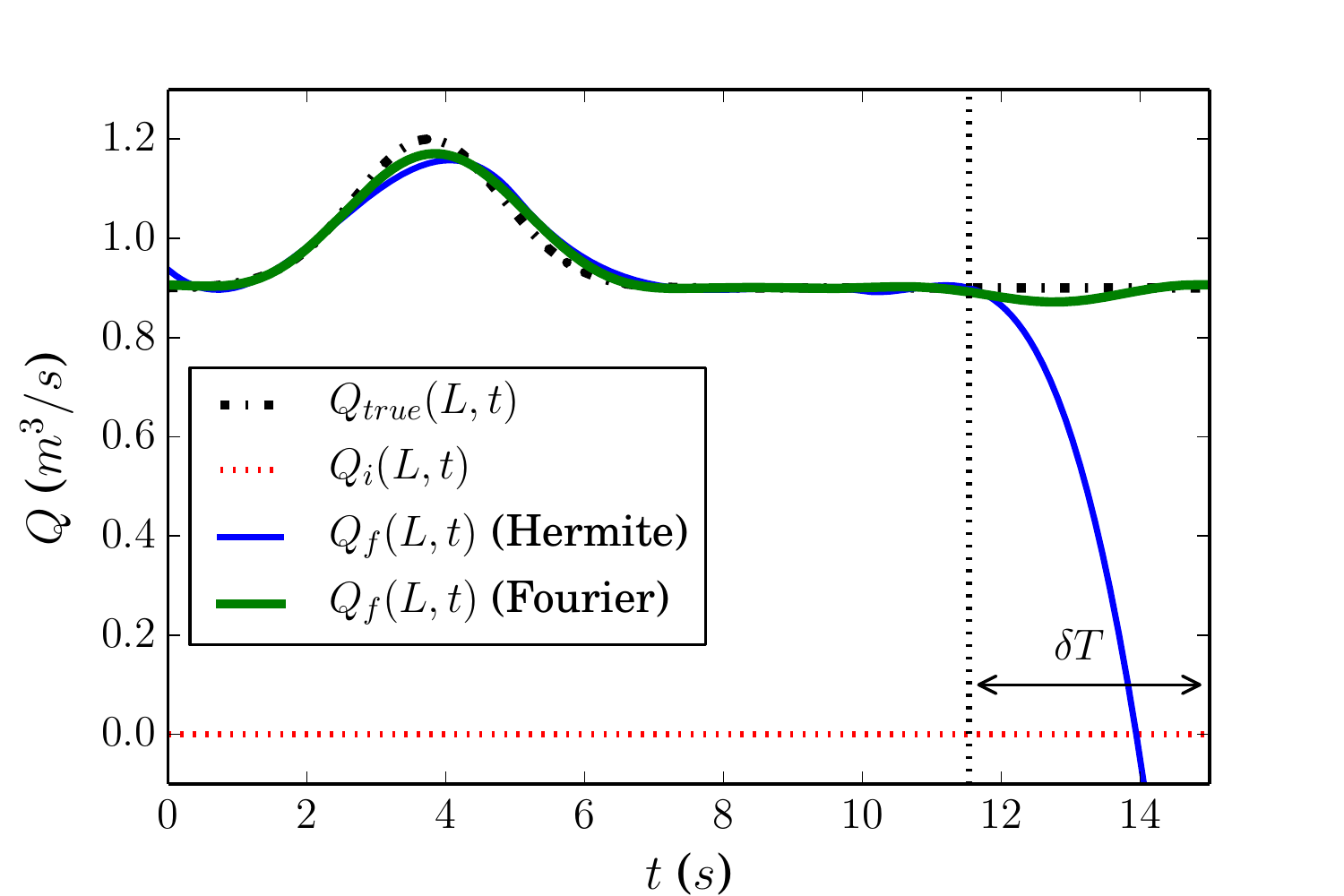} &
      \includegraphics[width=0.465\textwidth]{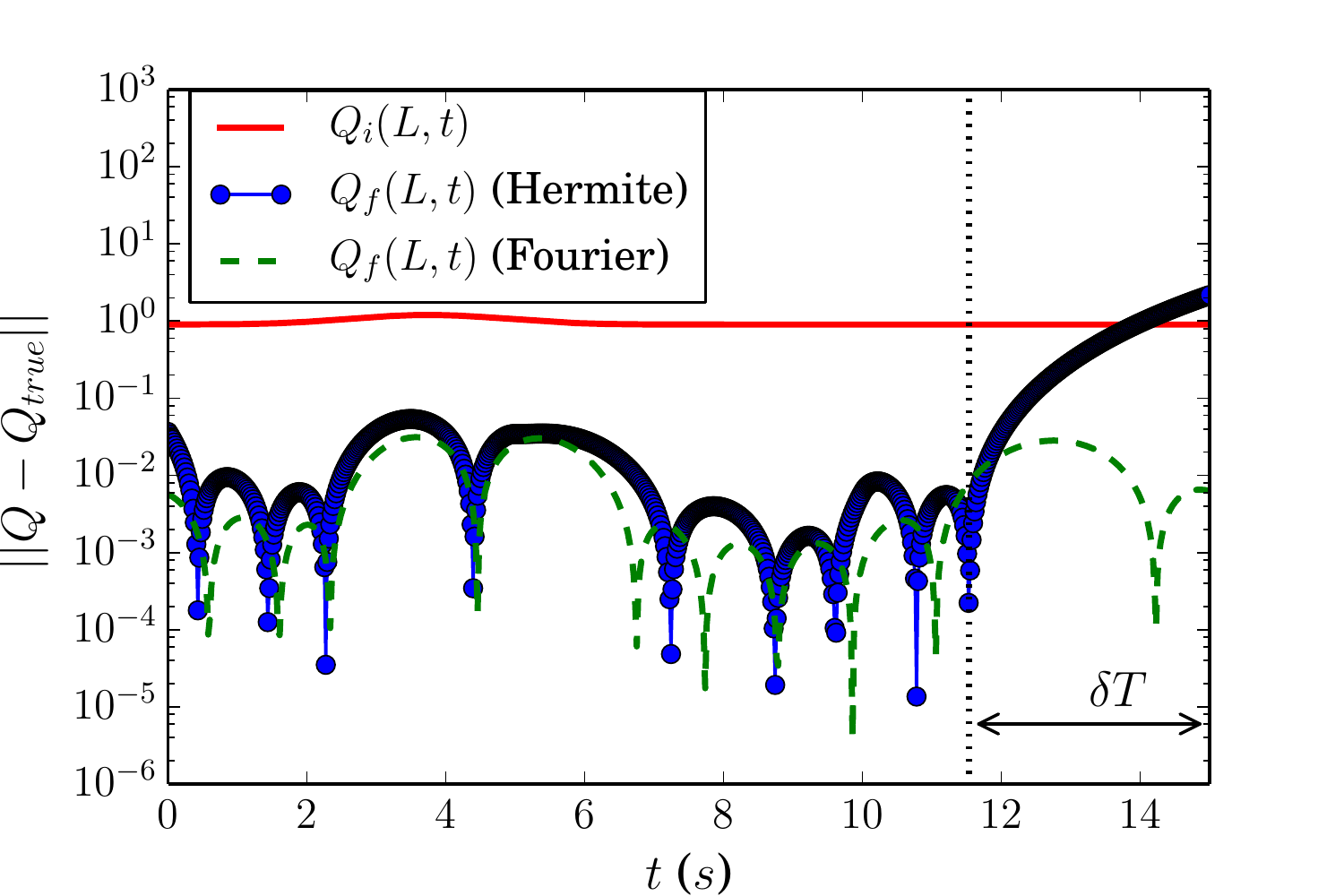} \\
      (a) & (b)
    \end{tabular}
  \end{center}
      \vspace{-2mm} 
  \caption{(a) Recovered time series $Q_f$ compared with $Q_\text{true}$ and
  $Q_i$. (b) Error \smash{$||Q(x,t_i)-Q_{true}(x,t_i)||$} at time $t_i$, where $||\cdot||_2$ is the discrete $L_2$ norm over the spatial domain. \label{fig:results_h}}
\end{figure}

We initialized the optimization routine with $Q_i(L,t)\equiv 0$. The time series $H^*(t)$ at the sensor resulting from the true boundary condition $Q_\text{true}$ and the initial proposed boundary condition $Q_i$ are shown in Figure \ref{fig:whargarbl}(b). The optimized time series $Q_f$ found with Hermite and Fourier representations are pictured against the initializing and true time series in Figure \ref{fig:results_h}(a). The run times, ratio of initial objective function value $f_i$ to final value $f_f$, and error in $Q(L,t)$ before and after optimization are shown in Table \ref{tab:tab3}. We considered both Hermite splines and Fourier modes with 8 degrees of freedom. Results are shown in Figure \ref{fig:results_h}. Notice that after time $\delta T = a/\delta x$, information has not had time to propagate from the boundary to the sensor so the Hermite representation has difficulty thereafter. The Fourier series representation maintains a low level of error through the whole time series as the true solution has the same value at $t=0$ as at $t=T$.

\begin{table}
  \caption{Results for boundary data recovery.\label{tab:tab3}}
      \vspace{-2mm} 
  \begin{center}
    \footnotesize
    \begin{tabular}{|l|l|l|}
      \hline
      & Hermite & Fourier \\
       \hline
CPU time (s)   &  176 &135\\ 
wall clock time (s)&  64& 51 \\ 
parallel speedup& 2.8& 2.6\\ 
$f_i$            &   $6.9\times10^4$& $6.9\times10^4$\\  
$f_f$            &   $1.9\times10^{-1}$& $3.2\times10^{-2}$\\   
$f_i/f_f$         &  $2.7\times10^{-6}$ &$4.6\times10^{-7}$ \\   
$||Q_i - Q_{true}||$ & 22.6& 22.6\\
$||Q_f - Q_{true}||$ & 0.477 &0.296\\
      \hline
    \end{tabular}
  \end{center}
\end{table}

\subsection{Reduction of variation in pressure}
\label{sec:opti2}
We next consider controlling boundary conditions to reduce potential damage to pipes in the water distribution system. Although the causes of pipe damage are complex and myriad, pressure transients play a role~\cite{Debon2010, Boulos2005}. Internal pressure contributes to axial, longitudinal and hoop stress in pipes~\cite{Rajani2001}, and pressure transients may help cause both pipe bursts (associated with high-pressure events) and collapses (associated low-pressure events)~\cite{Wang2013pipe}. During filling, the presence of trapped air may amplify pressure peaks and further increase damage~\cite{Vasconcelos2008, Kumpel2014}. Within the scope of the Preissman slot model, which does not explicitly describe air entrainment, we consider the potential for pipe damage by measuring, at each time step $t_i$, an estimated total variation in the pressure head $H$, which we denote $\langle dH/dx \rangle_i$, defined as
 \begin{equation}
\langle dH/dx \rangle_i  = \sum_{k=1}^{K}\sum_{j=0}^{N_{k}-1} |H^i_{k,j+1}-H^{i}_{k,j}|, \label{eqndh}
\end{equation}
where $K$ denotes the number of pipes in the network, $N_k$ the number of grid cells in pipe $k$, $M$ the number of time slices, and $H^i_{k,j} = H(A(x_j, t_i))$ for $A$ in pipe $k$. Note that
\begin{equation}
\langle dH/dx\rangle\approx \int_0^{L_i} \left|\frac{\partial H}{\partial x} \right| dx
\end{equation}
where $\partial H/\partial x$ may only exist in a weak sense due to the presence of shocks. The quantity $\langle dH/dx\rangle$ serves as proxy for pressure gradient that penalizes large water hammers associated with both high- and low- pressure events, and its minimization allows for exploration of operational regimes that potentially alleviate hydraulic contribution to pipe damage.

\begin{figure}
  \begin{center}
    \includegraphics[trim={2cm 0 2cm 1cm},clip,width = .9\textwidth]{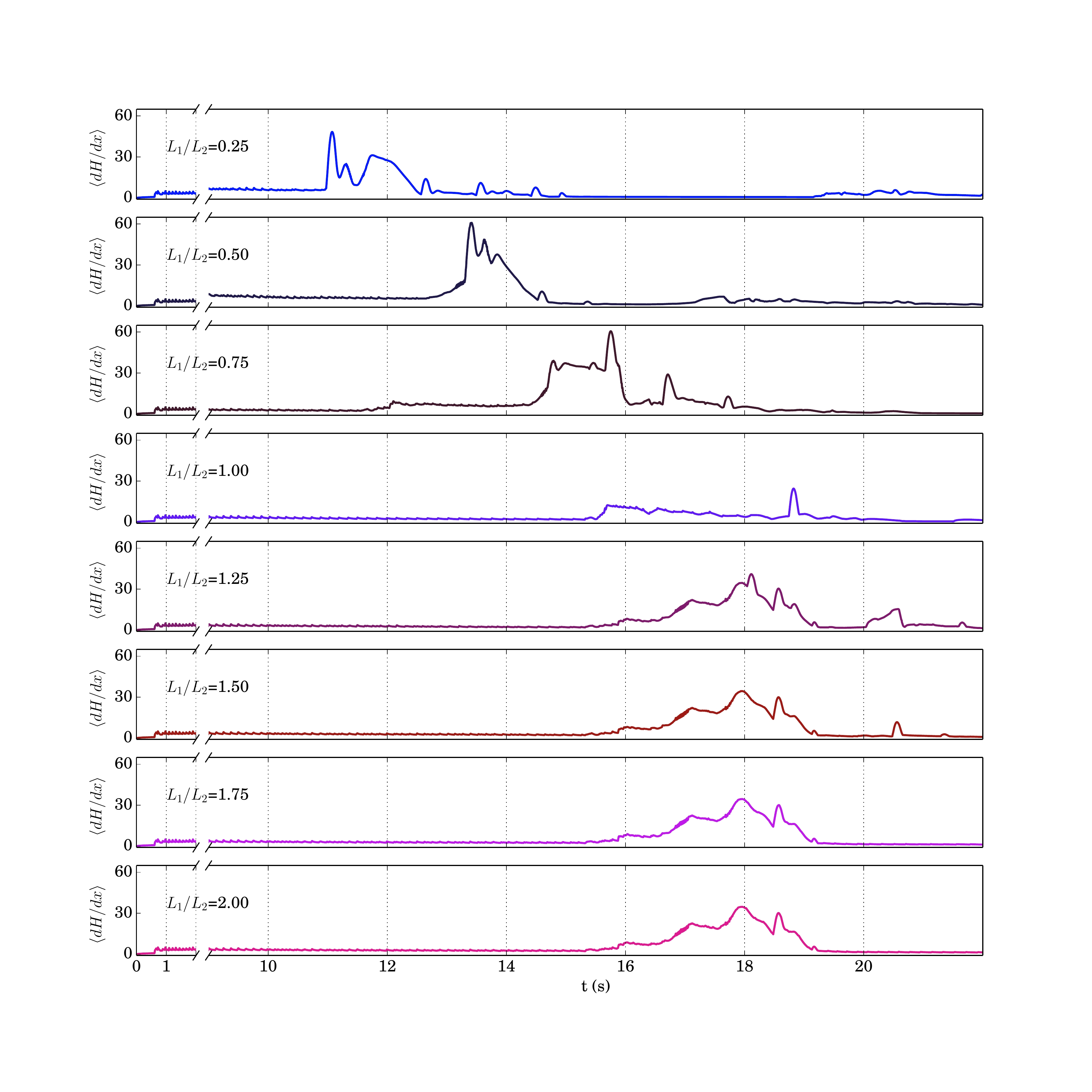}
  \end{center}
  \vspace{-1.2cm}
  \caption{Time series of $\langle dH/dx\rangle $ for different $L_1/L_2$.\label{fig:pretty2}}
\end{figure}

As we are interested in studying existing networks, rather than planning new ones, we first study how network geometry affects the time evolution of $\langle dH/dx \rangle$. We use the network structure shown in Figure \ref{fig:diag4} to study how the relative lengths of pipes in the network affects the interaction of reflected waves coming through a triple junction. We fix $L_0 = L_1 = 100\text{~m}$, and $L_2$ is varied between 25~m and 200~m. All boundaries have $Q=0$, and the initial conditions are $h = 0.8$~m in all pipes and $Q = 2$~m$^3$/s in pipe 0, and $Q=1$~m$^3$/s in pipes 1 and 2. These conditions were chosen to give a scenario where the flow transitions from free-surface to pressurized. The simulation time is 18~s and the pressure wave speed is 120~m/s. A time series of $\langle dH/dx\rangle$ for each network geometry is shown in Figure \ref{fig:pretty2}.

Mean and maximum values are tabulated in Table \ref{tab:dhdxp}. Interestingly, for $L_1=L_2$ we see the least dramatic pressure variation. Although the peak value of $\langle dH/dx \rangle$ varies by up to a factor of two over these different geometries, the mean values are quite comparable, suggesting that boundary control may still be useful even in different network geometries where reflected waves interact differently.

\begin{table}
  \caption{Mean and maximum $\langle dH/dx\rangle $ for different $L_1/L_2$. \label{tab:dhdxp}}
      \vspace{-2mm} 
  \begin{center}
    \footnotesize
    \begin{tabular}{|l|l|l|l|}
      \hline
      $L_1/L_2$ & Mean $\langle dH/dx\rangle $ & Max $\langle dH/dx\rangle $\\
      \hline
0.25    & 5.0609    &   48.3832 \\
0.50    & 5.4465    &   60.9356 \\
0.75    & 6.1335    &   60.6128 \\
1.00    & 3.4518    &   24.4900 \\
1.25    & 5.5090    &   40.9716 \\
1.50    & 5.1254    &   34.4893 \\
1.75    & 5.1019    &   34.5001 \\
2.00    & 5.1504    &   34.6804 \\
      \hline
    \end{tabular}
  \end{center}
\end{table}

\begin{figure}
  \begin{center}
    \begin{tabular}{c}
      \includegraphics[width=0.85\textwidth]{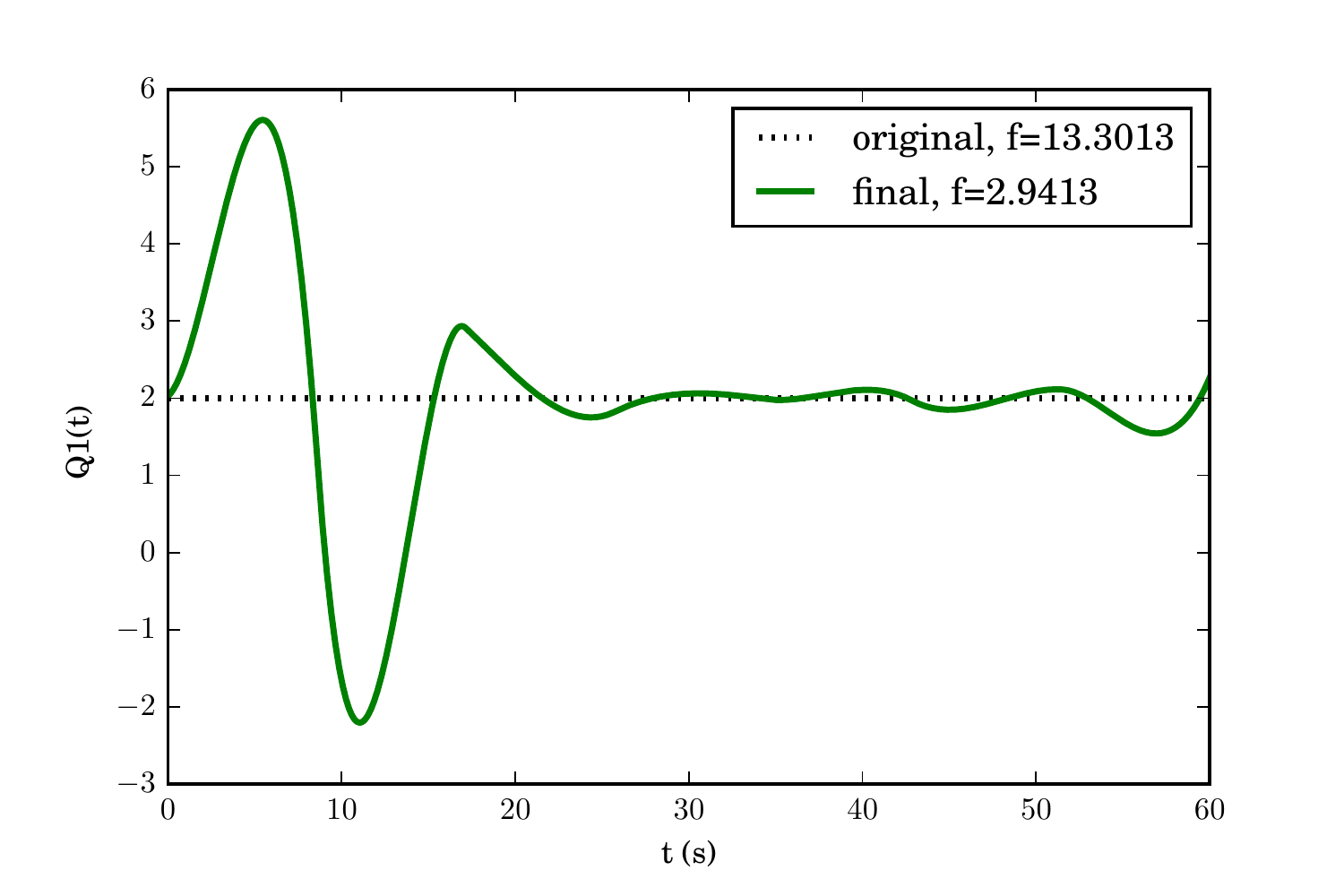} \\
      \includegraphics[width=0.85\textwidth]{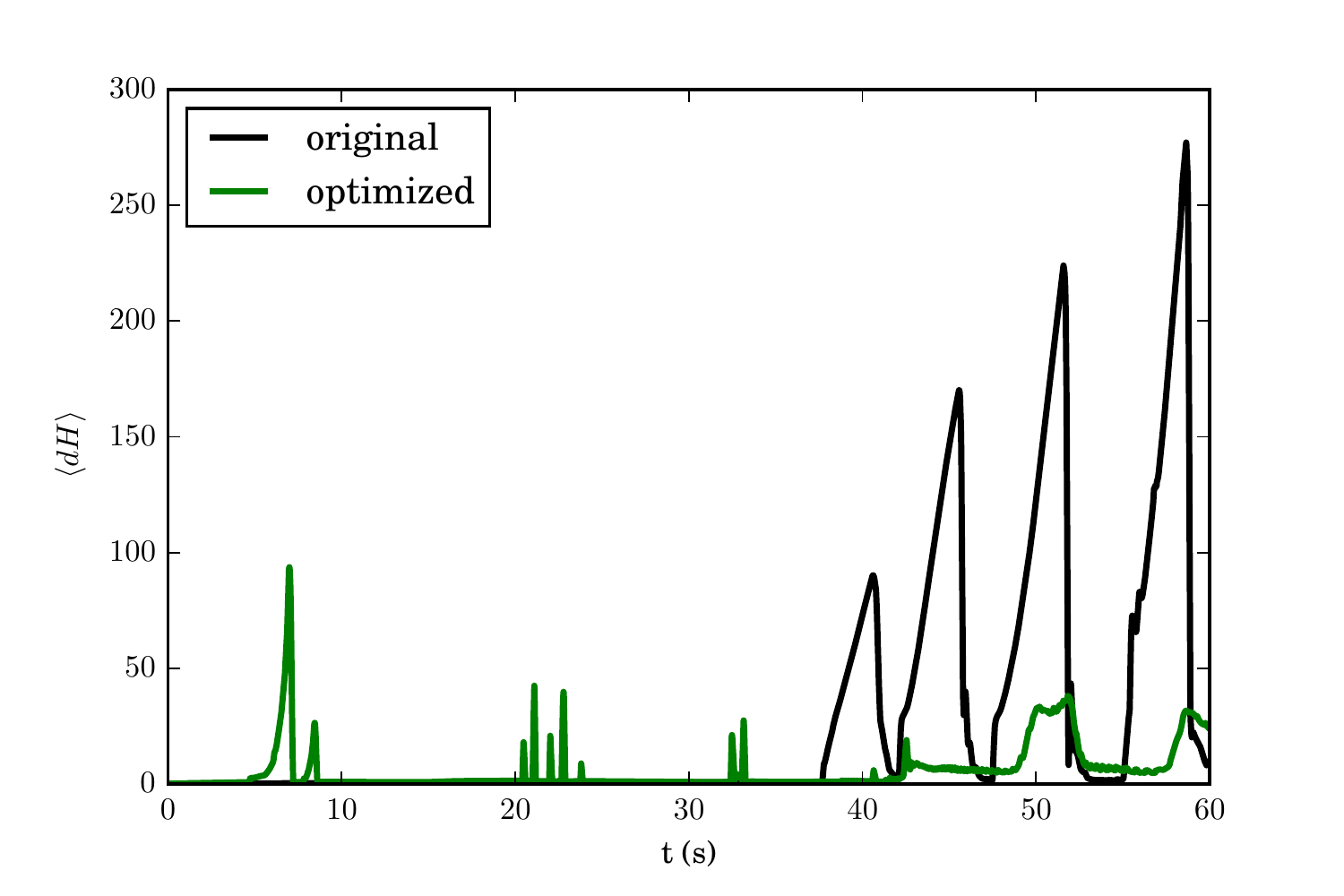} \\
    \end{tabular}
  \end{center}
    \vspace{-2mm} 
  \caption{Case (I) results. Top: Initializing and optimal time series $Q(0,t)$.
  Bottom: $\langle dH/dx\rangle$ over time for both inflow time
  series.\label{fig:optdh}}
\end{figure}

\begin{figure}
  \begin{center}
    \begin{tabular}{c}
      \includegraphics[width=0.85\textwidth]{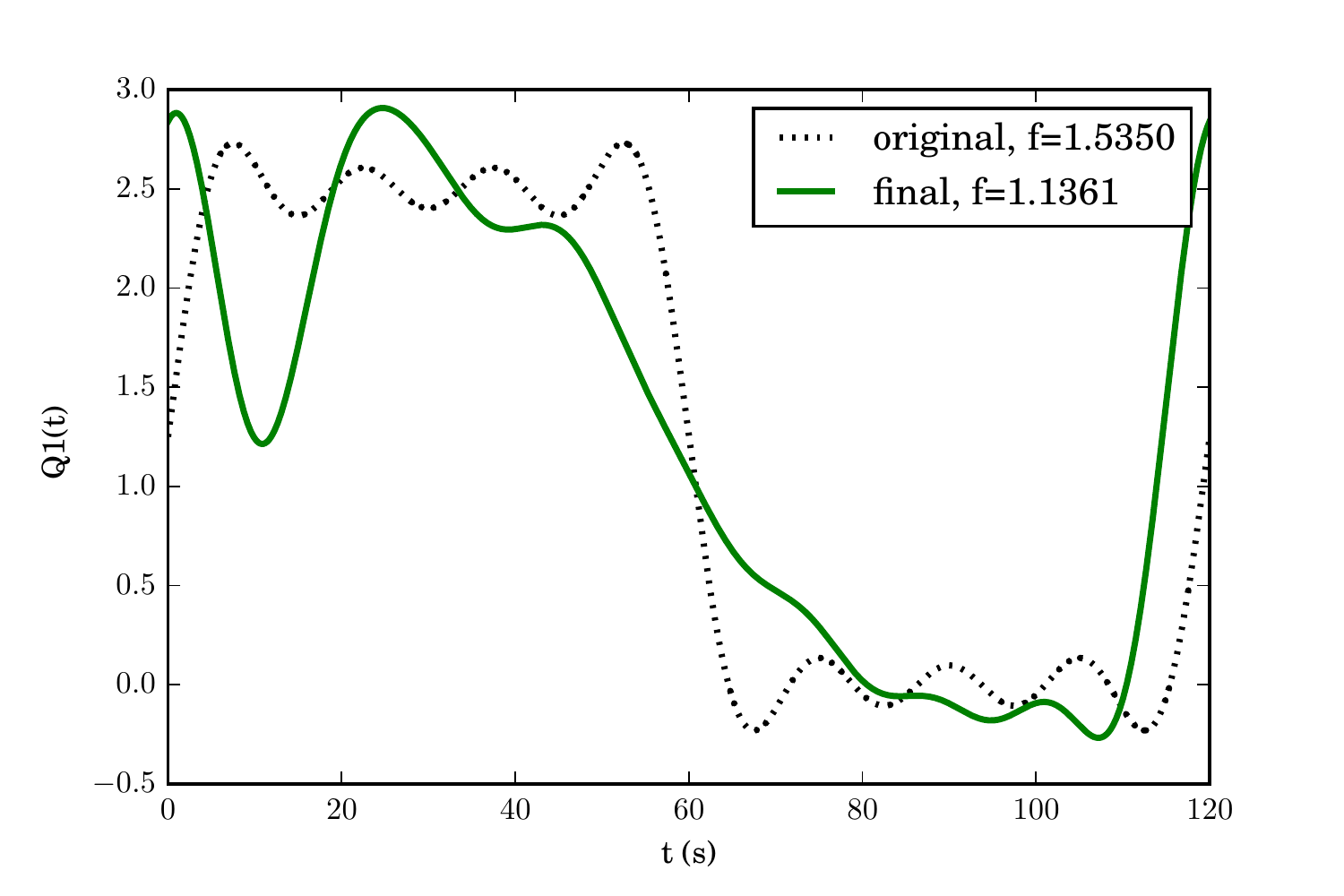} \\
      \includegraphics[width=0.85\textwidth]{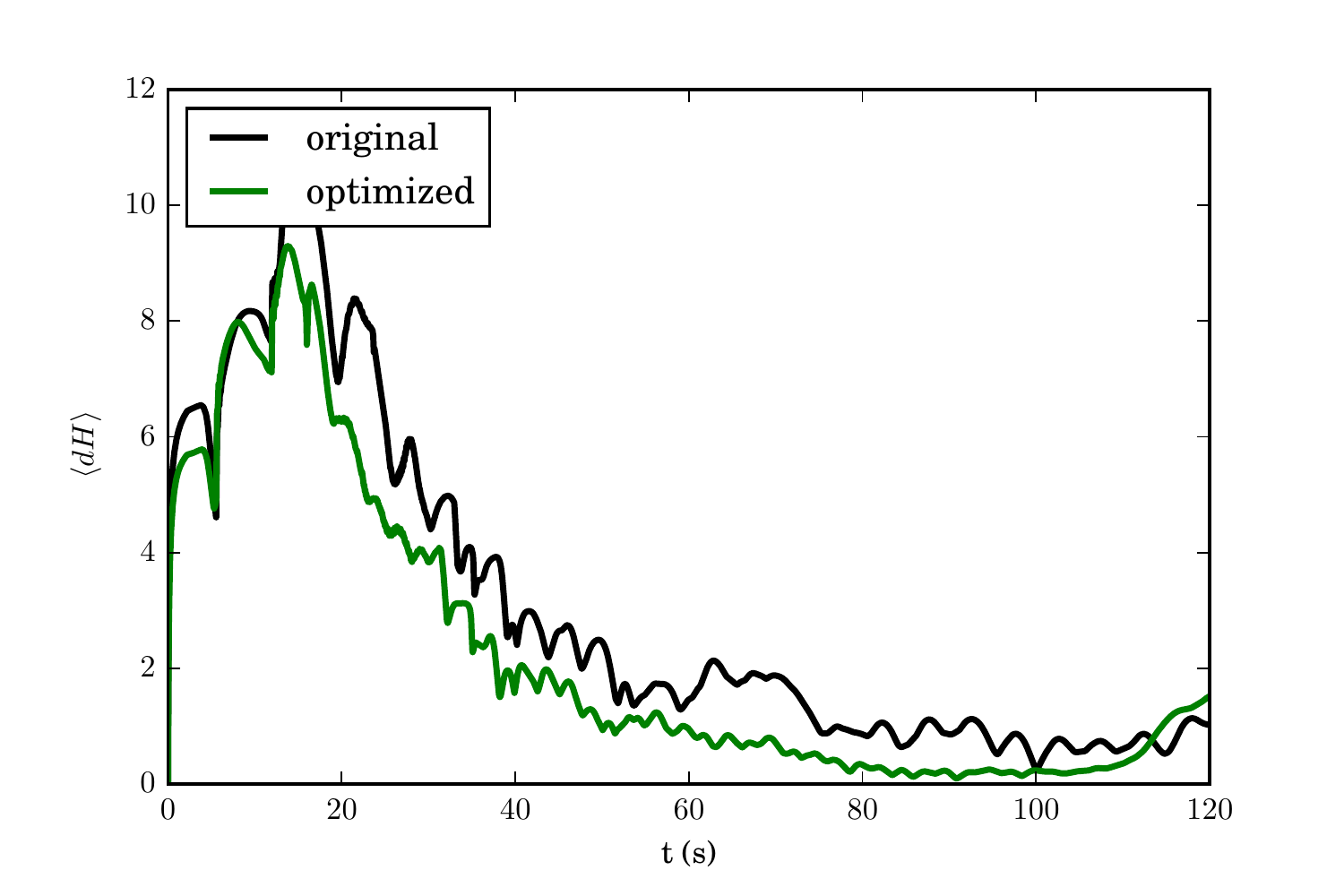} \\
    \end{tabular}
  \end{center}
    \vspace{-2mm} 
  \caption{Case (II) results. Top: Initializing and optimal time series
  $Q(0,t)$. Bottom: $\langle dH/dx\rangle$ over time for both inflow time
  series. \label{fig:optdj}}
\end{figure}

We now use the optimization framework to suggest inflow patterns minimizing $\langle dH/dx\rangle$ for one of these networks under a pressurization scenario. We define the objective function
\smash{$f=\frac{1}{2} ||\vec{r}||^2$}, where the components of $\vec{r}$ are given by
\begin{equation}
  r_i = \sqrt{\frac{1}{M}\langle dH/dx \rangle_i}
\end{equation}
for $i = 1,\ldots, M$. Motivated by practical applications, we suppose we want to supply a fixed volume of water $V_\ti$ over a time period $T$, and vary the inflow time series $Q(0,t)$ at the left end of pipe 0 in order to minimize $f$. Below we present three results highlighting both the usefulness and limitations of the framework. In all cases, the network connectivity is kept the same, and boundaries for the outflow ends of pipes 1 and 2 are set to $Q=0$. The initial conditions, time series representation, and simulation time are varied as follows:
\begin{romannum}
  \item[(I)] $L_0=L_1=100\text{~m}$, and $L_2=25\text{~m}$, roughness coefficient~=~0.015. Initial condition is all pipes empty. Degrees of freedom are cubic Hermite spline coefficients, with the first point $Q(0,0)$ determined by setting the integral of the Hermite time series equal to the desired inflow $V_\ti$. The initial constant inflow is compared with the optimized time series in Figure \ref{fig:optdh}(a). How $\langle dH/dx\rangle$ varies in time is shown in Figure \ref{fig:optdh}(b).
  \item[(II)] $L_0=L_1=100\text{~m}$, and $L_2=50\text{~m}$, roughness coefficient~=~0.008. Initial condition is $Q(x,t) = 5\text{~m/s}$ in pipe 0, with pipes 1 and 2 empty. The degrees of freedom are Fourier modes, with the constant mode constrained to $2V_\ti/T$, such that $V_\ti$ is the total inflow volume. The optimization starts with a time series $Q(0,t)$ for pipe 0 which approximates a step function inflow. Pressure wave speed $a=10\text{~m/s}$ and simulation time $T=120\text{~s}$. Results are pictured in Figure \ref{fig:optdj}.
  \item[(III)] As in case (II), but with pressure wave speed $a=100\text{~m/s}$ and $T=16\text{~s}$.
\end{romannum}
Results for case (II) were computed with a relatively wide slot width, corresponding to a gravity wave speed of $a = 10\text{~m/s}$. With a more realistic pressure wave speed of 100~m/s in case (III), the algorithm found only very slight improvements for the inflow time series, as shown in the table above. This is probably due to the presence of unphysical oscillations accompanying the initial pressurization event. That the Preissman slot model, with a narrow slot width, gives rise to such oscillations during rapid pressurization has been noted previously by Vasconcelos~\cite{Vasconcelos2009}, among others. Vasconcelos suggests filtering or artificial viscosity to damp out these fluctuations, but also concedes the averaged behavior may present a reasonable approximation of reality. Even if these oscillations are filtered, averaged, or damped into submission, they limit the utility of the Preissman slot model for informing smooth optimization when rapid pressure transients are present. However, in regimes with more gradual filling or emptying, the Preissman slot model may suffice for both pressure gradient optimization and boundary value recovery.

\begin{table}
  \caption{Numerical results for pressure gradient reduction.\label{tab:tab4}}
    \vspace{-2mm} 
  \begin{center}
    \footnotesize
    \begin{tabular}{|l| l| l| l| l| l| l| l| l| l| l|}
      \hline
      Case & DOF & $a$& T &M & CPU time & Wall time & $ f_0$ & $f_f$/$f_0$ & $ V_\ti$\\
      \hline
      I& 15 (Hermite) &  100 & 60 &15000&    5298&  772&               13.3013&   0.2211  & 120\\
       II & 15 (Fourier) & 10 &   120& 4000&     2819&     405&     1.53498 &     0.7401  & 150\\
      III & 15 (Fourier) &100&  16&   3600&     2576&    350  &    1.83238&    0.9988  &   20\\
      \hline
    \end{tabular}
  \end{center}
\end{table}

\section{Conclusions and Future Work}
In this work we have introduced the study and management of intermittent water supply as a problem in applied mathematics, a problem with interesting theoretical facets as well as compelling practical implications. We present results from a combined modeling and optimization framework that captures some of the important dynamics in the systems of interest and demonstrates two optimization examples with real-world significance. Although the present results may help guide some ideas for managing pressurized flow, model improvements and alternative optimization schemes may be necessary to address applications of pressure gradient reduction and other objectives in larger and/or poorly-mapped networks.

In the future we hope to more carefully extract the physics and timescales relevant to the application of intermittent water supply in a robust model that retains the computational tractability of the current one. The code allows for straightforward implementation of alternative models for single-pipe flow and boundary coupling, allowing us to make the pressure gradient reduction optimizations more relevant to real-world data. We also hope deal with questions of water quality and to extend our implementation to tackle larger networks and other objective functions, aspiring to examine, understand, and perhaps improve real-world scenarios.

\section*{Acknowledgments}
We thank John Erickson and Kara Nelson (UC Berkeley Civil Engineering) for their insight and network layout data.

\appendix
\section{Fractional-power scaled Chebyshev approximation}
\label{app:cheby}
The numerical method frequently requires the height $h$ as a function of the cross-sectional area $A$. In the Preissman slot geometry, there is no analytic expression for $h(A)$ below the pipe crown, necessitating some form of approximation. In what follows, all variables are assumed to be non-dimensionalized, e.g. $A\rightarrow A/D^2$ and $h\rightarrow h/D$.

One strategy is to use the relation \smash{$A=\frac{1}{8}(\theta-\sin(\theta))$} to solve for $\theta$, and evaluate \smash{$h = \frac{1}{2}(1-\cos(\theta/2))$}. However, rootfinding is cumbersome as the computation must be performed for every cell during every time step. Furthermore, the authors found that representation of $h(A)$ with a standard Chebyshev or other polynomial interpolant proved strikingly useless due to singular behavior at the origin. Comparing Taylor series shows that \smash{$h \sim A^{2/3}$} near $h=0$. Using a Chebyshev interpolant of the form
\begin{equation}
  h(A) = \sum_{k=0}^N a_k T_k\left(cA^{2/3}-1\right),
\end{equation}
where $T_k$ denotes the $k^\text{th}$ Chebyshev polynomial on $[-1,1]$, scales out the singular behavior and ensures accuracy. The Chebyshev representation allows for fast evaluation via a three-term recursion. The current version of the code uses $N = 20$ Chebyshev points for $h(A)$, a change which speeds up the calculation of $h(A)$ by a factor of 7 when compared to root-bracketing, without sacrificing double-precision accuracy. (Newton's method is not desirable for this computation, as the derivative of \smash{$f(\theta) = A-\frac{1}{8}(\theta-\sin(\theta))$} is zero at the origin).

We used the same strategy to better approximate the quantity $\phi(A)$, which shows up in the speed estimation for the HLL solver and cannot be analytically expressed in terms of $A$. Several other authors use a ninth order McLaurin series for expressing $\phi$ as a function of $\theta$~\cite{Leon2006,Kerger2011}, which avoids repeated quadrature but still relies on root finding for $\theta$ in terms of $A$. Furthermore, this expression also loses accuracy at the right end of the interval, with errors of order $10^{-2}$ near the pipe crown. 
 
\begin{figure}
  \begin{center}
    \begin{tabular}{c}
      \includegraphics[width=0.9\textwidth]{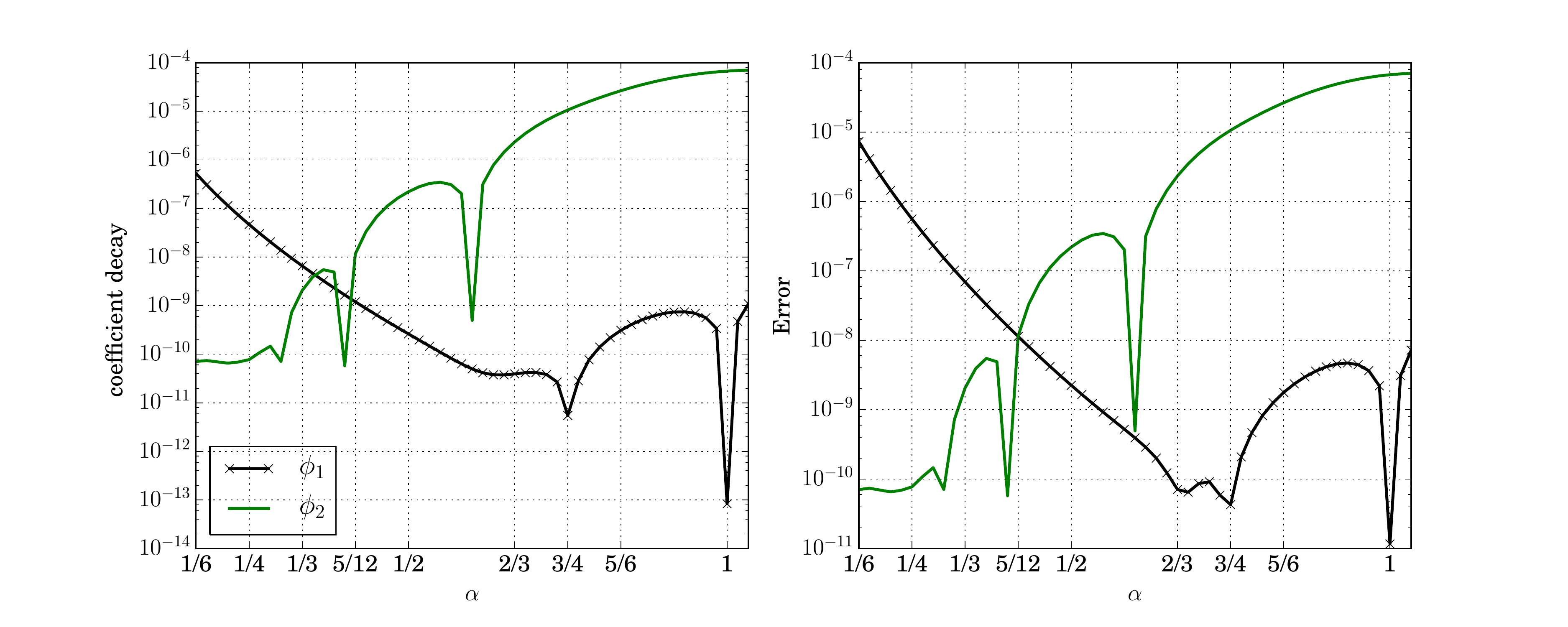}\\
      \includegraphics[width=0.9\textwidth]{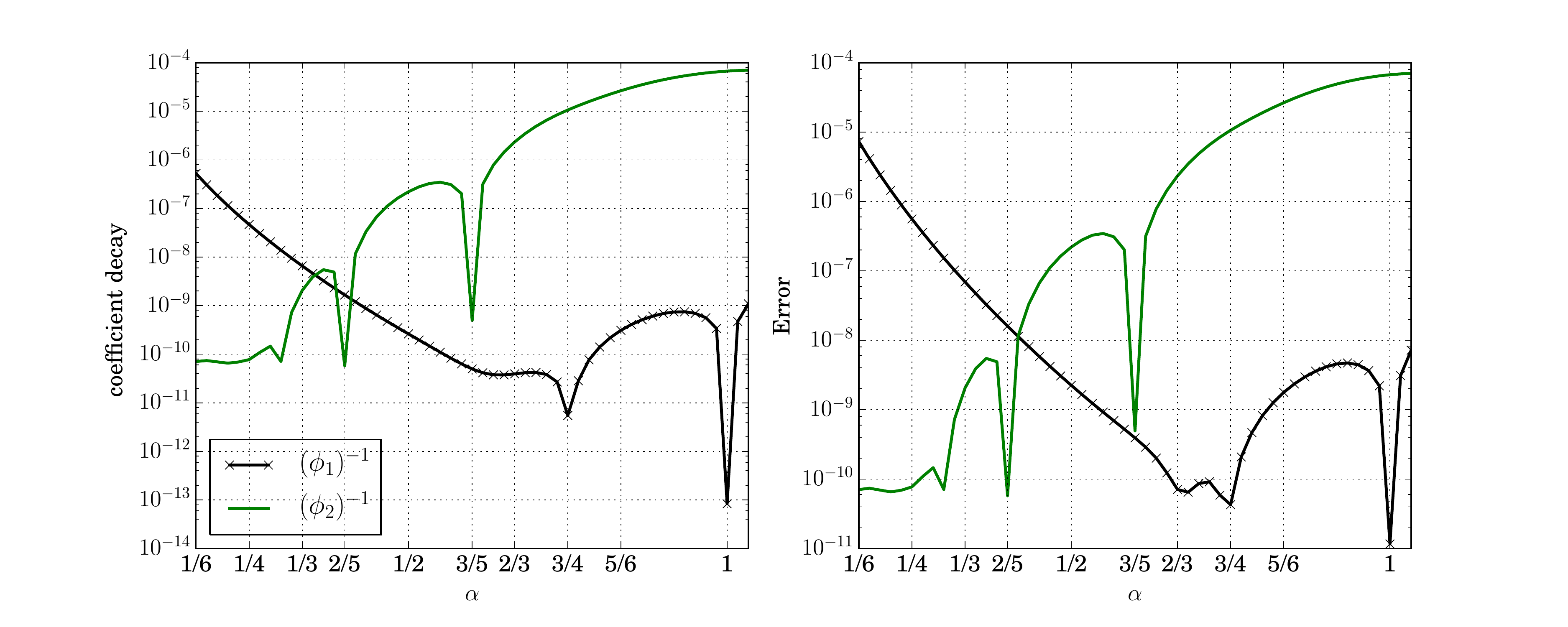}
    \end{tabular}
  \end{center}
  \vspace{-5mm} 
  \caption{Coefficient decay and error as a function of power scaling $\alpha$
  with $N=20$ terms, for $\phi_i(A)$ (top) and $A_i$ (bottom). The black
  hatched lines denote the expansion on the first half of interval ($i=1$), and
  the green line denotes the expansion on the second half ($i=2$).\label{fig:ff}}
\end{figure}

The authors thus set out to improve both speed and accuracy in computation of $\phi$ by finding a Chebyshev representation for both $\phi(A)$ and $\phi^{-1} = A(\phi)$, of the form
\begin{equation}
\phi(A) = \sum_{k=0}^{N} a_k T_k(cA^{\alpha}-1), \quad A(\phi) = \sum_{k=0}^N a_k T_k(c\phi^{\alpha}-1).
\end{equation}
As this geometry has peculiarities near both the pipe floor and pipe crown, we considered separate expansions on the first and second halves of the interval of interest. By numerical experiment with accuracy and coefficient decay of expansions, we found the best values of $\alpha$ for four expansions summarized in Table \ref{tab:te}. Figure \ref{fig:ff} shows the accuracy and coefficient decay dependence on $\alpha$ for each of these cases. Our implementation uses $N=20$ terms in the expansions, but we tried several other values of $N$ during our numerical experiments to make sure that that the number of Chebyshev points did not affect the scaling choices. 

\begin{table}
  \caption{Scaling parameter $\alpha$ values chosen.\label{tab:te}}
    \vspace{-2mm} 
  \begin{center}
    \footnotesize
    \begin{tabular}{|l|l|l|}
      \hline
      Function & Range& $\alpha$\\
      \hline
      $\phi_1(A)$ & $0\leq A<\pi/8$& 1/3\\
      \hline
      $\phi_2(A)$ & $\pi/8\leq A\leq A_t$ & 5/12\\
      \hline
      $A_1(\phi)$   & $0\leq \phi< \phi(\pi/8)$ & 1\\
      \hline
      $A_2(\phi) $  & $\phi(\pi/8)<\phi<\phi(A_t)$ &3/5\\
      \hline
    \end{tabular}
  \end{center}
\end{table}
\vspace{-1cm}

\newpage

\bibliographystyle{siam}

\begin{thebibliography}{10}

\bibitem{Borsche2014}
{\sc R.~Borsche and A.~Klar}, {\em {Flooding in urban drainage systems:
  coupling hyperbolic conservation laws for sewer systems and surface flow}},
  {Internat. J. Numer. Methods Fluids}, {76} ({2014}), pp.~{789--810}.

\bibitem{Boulos2005}
{\sc P.~F. Boulos, B.~W. Karney, D.~J. Wood, and S.~Lingireddy}, {\em Hydraulic
  transient guidelines for protecting water distribution systems}, J Am Water
  Works Ass),  (2005), pp.~111--124.

\bibitem{Bourdarias2007a}
{\sc C.~Bourdarias and S.~Gerbi}, {\em {A finite volume scheme for a model
  coupling free surface and pressurised flows in pipes}}, {J. Comput. Appl.
  Math.}, {209} ({2007}), pp.~{109--131}.

\bibitem{Bourdarias2008}
{\sc C.~Bourdarias, S.~Gerbi, and M.~Gisclon}, {\em {A kinetic formulation for
  a model coupling free surface and pressurised flows in closed pipes}}, {J.
  Comput. Appl. Math.}, {218} ({2008}), pp.~{522--531}.

\bibitem{Bousso2013}
{\sc S.~Bousso, M.~Daynou, and M.~Fuamba}, {\em {Numerical Modeling of Mixed
  Flows in Storm Water Systems: Critical Review of Literature}}, {J. Hydraul.
  Eng.-ASCE}, {139} ({2013}), pp.~{385--396}.

\bibitem{Capart1999}
{\sc H.~Capart, C.~Bogaerts, J.~Kevers-Leclercq, and Y.~Zech}, {\em {Robust
  numerical treatment of flow transitions at drainage pipe boundaries}}, {Water
  Sci. Technol.}, {39} ({1999}), pp.~{113--120}.
\newblock {4th International Conference on Developments in Urban Drainage
  Modelling (UDM 98), London, England, SEP 21--24, 1998}.

\bibitem{Christodoulou2012}
{\sc S.~Christodoulou and A.~Agathokleous}, {\em {A study on the effects of
  intermittent water supply on the vulnerability of urban water distribution
  networks}}, {Water Sci Technol-Water Supply}, {12} ({2012}), pp.~{523--530}.

\bibitem{Colombo2008}
{\sc R.~M. Colombo, M.~Herty, and V.~Sachers}, {\em On $2\times2$ conservation
  laws at a junction}, SIAM J. Math. Anal., 40 (2008), pp.~605--622.

\bibitem{DeMarchis2010}
{\sc M.~De~Marchis, C.~M. Fontanazza, G.~Freni, G.~La~Loggia, E.~Napoli, and
  V.~Notaro}, {\em {A model of the filling process of an intermittent
  distribution network}}, {Urban Water Journal}, {7} ({2010}), pp.~{321--333}.

\bibitem{DeMarchis2013}
{\sc M.~De~Marchis, C.~M. Fontanazza, G.~Freni, G.~La~Loggia, V.~Notaro, and
  V.~Puleo}, {\em {A mathematical model to evaluate apparent losses due to
  meter under-registration in intermittent water distribution networks}},
  {Water Sci Technol-Water Supply}, {13} ({2013}), pp.~{914--923}.

\bibitem{Debon2010}
{\sc A.~Deb\'{o}n, A.~Carrion, E.~Cabrera, and H.~Solano}, {\em {Comparing risk
  of failure models in water supply networks using ROC curves}}, {Rel. Eng.
  Sys. Safety}, {95} ({2010}), pp.~{43--48}.

\bibitem{Freni2014}
{\sc G.~Freni, M.~De~Marchis, and E.~Napoli}, {\em {Implementation of pressure
  reduction valves in a dynamic water distribution numerical model to control
  the inequality in water supply}}, {Journal of Hydroinformatics}, {16}
  ({2014}), pp.~{207--217}.

\bibitem{Ghidauoi2005}
{\sc M.~Ghidaoui, {Ming Zhao}, D.~McInnis, and D.~Axworthy}, {\em {A review of
  water hammer theory and practice}}, {Applied Mechanics Review}, {58}
  ({2005}), pp.~{49--76}.

\bibitem{Gottlieb1998}
{\sc S.~Gottlieb and C.~Shu}, {\em {Total variation diminishing Runge-Kutta
  schemes}}, {Mathematics of Computation}, {67} ({1998}), pp.~{73--85}.

\bibitem{Kerger2012}
{\sc F.~Kerger, P.~Archambeau, B.~J. Dewals, S.~Erpicum, and M.~Pirotton}, {\em
  {Three-phase bi-layer model for simulating mixed flows}}, {J. Hydraul. Res.},
  {50} ({2012}), pp.~{312--319}.

\bibitem{Kerger2011}
{\sc F.~Kerger, P.~Archambeau, S.~Erpicum, B.~J. Dewals, and M.~Pirotton}, {\em
  {An exact Riemann solver and a Godunov scheme for simulating highly transient
  mixed flows}}, J. Comput. Appl. Math., 235 (2011), pp.~2030--2040.

\bibitem{kumpel2013}
{\sc E.~Kumpel and K.~L. Nelson}, {\em Comparing microbial water quality in an
  intermittent and continuous piped water supply}, Water Research, 47 (2013),
  pp.~5176--5188.

\bibitem{Kumpel2014}
{\sc E.~Kumpel and K.~L. Nelson}, {\em Mechanisms affecting water quality in an
  intermittent piped water supply}, Environmental Science and Technology, 48
  (2014), pp.~2766--2775.

\bibitem{Leon2010a}
{\sc A.~Le\'on, A.~R. Schmidt, P.~D, , M.~H. Garc\'{\i}a, and P.~a. D}, {\em
  {Junction and Drop-Shaft Boundary Conditions for Modeling Free-Surface ,
  Pressurized , and Mixed Free-Surface Pressurized Transient Flows}}, J.
  Hydraul. Eng.-ASCE, 136 (2010), pp.~705--715.

\bibitem{ITM}
{\sc A.~S. Leon}, {\em Illinois Transient Model: Two-Equation Model V. 1.3
  User's Manual}, 2011.
\newblock \url{http://web.engr.oregonstate.edu/~leon/ITM.htm}.

\bibitem{Leon2006}
{\sc A.~S. Le\'{o}n, M.~S. Ghidaoui, , A.~R. Schmidt, , and M.~H.~a.
  Garc\'{\i}a}, {\em {Godunov-Type Solutions for Transient Flows in Sewers}},
  J. Hydraul. Eng.-ASCE, 132 (2006), pp.~800--813.

\bibitem{Leon2009}
{\sc A.~S. Le\'{o}n, M.~S. Ghidaoui, A.~R. Schmidt, and M.~H. Garc\'{\i}a},
  {\em {Application of Godunov-type schemes to transient mixed flows}}, J.
  Hydraul. Res., 47 (2009), pp.~37--41.

\bibitem{Leon2010}
{\sc A.~S. Le\'on, M.~S. Ghidaoui, A.~R. Schmidt, and M.~H. Garc\'ia}, {\em {A
  robust two-equation model for transient-mixed flows}}, J. Hydraul. Res.,
  (2010), pp.~37--41.

\bibitem{Leveque2002}
{\sc R.~Le{V}que}, {\em Finite Volume Methods for Hyperbolic Problems},
  Cambridge University Press, 2002.

\bibitem{Preissman1961}
{\sc A.~Preissman}, {\em {Propagation des intumescences dans les canaux et
  rivieres}}, in First Congress of the French Association for Computation,
  Grenoble, France, 1961.

\bibitem{Rajani2001}
{\sc B.~Rajani and Y.~Kleiner}, {\em Comprehensive review of structural
  deterioration of water mains: physically based models}, Urban Water, 3
  (2001), pp.~151--164.

\bibitem{EPA}
{\sc L.~Rossman}, {\em EPANET 2 User Manual}, United States Environmental
  Protection Agency, 2000.

\bibitem{SWMM}
\leavevmode\vrule height 2pt depth -1.6pt width 23pt, {\em Storm Water
  Management Model User's Manual Version 5.0}, United States Environmental
  Protection Agency, 2010.

\bibitem{Sanders2000}
{\sc B.~Sanders and N.~Katopodes}, {\em {Adjoint sensitivity analysis for
  shallow-water wave control}}, J. Eng. Mech.-ASCE,  (2000), pp.~909--919.

\bibitem{Sanders2011}
{\sc B.~F. Sanders and S.~F. Bradford}, {\em {Network Implementation of the
  Two-Component Pressure Approach for Transient Flow in Storm Sewers}}, J.
  Hydraul. Eng.-ASCE, 137 (2011), pp.~158--172.

\bibitem{Sharma2009}
{\sc S.~K. Sharma and K.~Vairavamoorthy}, {\em {Urban water demand management :
  prospects and challenges for the developing countries}}, Water and
  Environment Journal, 23 (2009), pp.~210--218.

\bibitem{Trajkovic1999}
{\sc B.~Trajkovic, M.~Ivetic, F.~Calomino, and A.~D'Ippolito}, {\em
  {Investigation of transition for free surface to pressurized flow in a
  circular pipe}}, Water Sci. Technol., 39 (1999), pp.~105--112.

\bibitem{Vairavamoorthy2008}
{\sc K.~Vairavamoorthy, S.~D. Gorantiwar, and A.~Pathirana}, {\em {Managing
  urban water supplies in developing countries , Climate change and water
  scarcity scenarios}}, Physics and Chemistry of the Earth, 33 (2008),
  pp.~330--339.

\bibitem{Vasconcelos2011}
{\sc J.~G. Vasconcelos, A.~M. Asce, and D.~T.~B. Marwell}, {\em {Innovative
  Simulation of Unsteady Low-Pressure Flows in Water Mains}}, J. Hydraul.
  Eng.-ASCE, 137 (2011), pp.~1490--1499.

\bibitem{Vasconcelos2008}
{\sc J.~G. Vasconcelos and S.~J. Wright}, {\em {Rapid flow startup in filled
  horizontal pipelines}}, {J. Hydraul. Eng.-ASCE}, {134} ({2008}),
  pp.~{984--992}.

\bibitem{Vasconcelos2007}
{\sc J.~G. Vasconcelos, S.~J. Wright, D.~Ribeiro, P.~Sg, and A.~Norte}, {\em
  {Comparison between the two-component pressure approach and current transient
  flow solvers}}, J. Hydraul. Res., 45 (2007), pp.~37--41.

\bibitem{Vasconcelos2006}
{\sc J.~G. Vasconcelos, S.~J. Wright, and P.~L. Roe}, {\em {Improved Simulation
  of Flow Regime Transition in Sewers: Two-Component Pressure Approach}}, J.
  Hydraul. Eng.-ASCE, 132 (2006), pp.~553--562.

\bibitem{Vasconcelos2009}
{\sc J.~G. Vasconcelos, S.~J. Wright, and P.~L. Roe}, {\em {Numerical
  Oscillations in Pipe-Filling Bore Predictions by Shock-Capturing Models}},
  {J. Hydraul. Eng.-ASCE}, {135} ({2009}), pp.~{296--305}.

\bibitem{Wang2013pipe}
{\sc R.~Wang, Z.~Wang, X.~Wang, H.~Yang, and J.~Sun}, {\em Pipe burst risk
  state assessment and classification based on water hammer analysis for water
  supply networks}, Water Sci Technol, 140 (2013), p.~04014005.

\bibitem{Wylie1993}
{\sc E.~B. Wylie and V.~L. Streeter}, {\em Fluid Transients in Systems},
  Prentice Hall, 1993.

\end{thebibliography}

\end{document}